%% file: graphit.tex
\documentclass[10pt, conference]{IEEEtran}
\usepackage{algorithm}
\usepackage{algorithmicx}
\usepackage{algpseudocode}
\usepackage{color} 
\usepackage{graphics}
\usepackage{graphicx}
\usepackage[font={small},skip=6pt,belowskip=-10pt]{caption}
\usepackage[belowskip=2pt]{subcaption}
\usepackage{comment}
\let\labelindent\relax
\usepackage{enumitem}
\usepackage{xspace}

\newcommand\punt[1]{}

\newcommand{\segmenting}{CSR segmenting\xspace}
\newcommand{\Segmenting}{CSR Segmenting\xspace}
\newcommand{\Cagra}{Cagra\xspace}

\newcommand{\reordering}{frequency based clustering\xspace}
\newcommand{\Reordering}{Frequency Based Clustering\xspace}
\newcommand{\Baseline}{HandOpt C++\xspace}
\newcommand{\baseline}{hand optimized C++\xspace}

\begin{document}

\newenvironment{denseitemize}{
\begin{itemize}[topsep=2pt, partopsep=0pt, leftmargin=1.5em]
  \setlength{\itemsep}{4pt}
  \setlength{\parskip}{0pt}
  \setlength{\parsep}{0pt}
}{\end{itemize}}

%
\title{\Large \bf Making Caches Work for Graph Analytics}


\author{\IEEEauthorblockN{Yunming Zhang, Vladimir Kiriansky, \\ Charith Mendis, Saman Amarasinghe}
\IEEEauthorblockA{MIT CSAIL\\
\{yunming,vlk,charithm,saman\}@csail.mit.edu}
\and
\IEEEauthorblockN{Matei Zaharia}
\IEEEauthorblockA{Stanford InfoLab\\
matei@cs.stanford.edu}
}

\maketitle

\input{abstract}

%

\input{intro}

\input{understanding}

\input{segmenting}

\input{programming}

\input{reorder}

\input{evaluation}

\input{related}

\input{conclusion}
\input{ack}

%
\IEEEpeerreviewmaketitle

{\footnotesize \bibliographystyle{IEEEtran}
\bibliography{graph.bib}}

\end{document}

%% file: abstract.tex
\begin{abstract}
Large-scale applications implemented in today's high performance graph frameworks heavily underutilize modern hardware systems. While many graph frameworks have made substantial progress in optimizing these applications, we show that it is still possible to achieve up to 5$\times$ speedups over the fastest frameworks by greatly improving cache utilization. Previous systems have applied out-of-core processing techniques
  from the memory/disk boundary to the cache/DRAM boundary.  However,
  we find that blindly applying such techniques is ineffective because
  the much smaller performance gap between cache and DRAM requires new designs for achieving scalable performance and low overhead. We present \Cagra, a cache optimized in-memory graph framework. \Cagra uses a novel technique, \Segmenting, to break the vertices into segments that fit in last level cache, and partitions the graph into subgraphs based on the segments. Random accesses in each subgraph are limited to one segment at a time, eliminating the much slower random accesses to DRAM. The intermediate updates from each subgraph are written into buffers sequentially and later merged using a low overhead parallel cache-aware merge. \Cagra achieves speedups of up to 5$\times$ for PageRank, Collaborative Filtering, Label Propagation and Betweenness Centrality over the best published results from state-of-the-art graph frameworks, including GraphMat, Ligra and GridGraph.

\end{abstract}


%% file: intro.tex
\section{Introduction}
\label{sec:intro}



High performance graph analytics has received considerable research attention, leading to a series of optimized frameworks such as GraphLab~\cite{low12vldb-distr-graphlab}, Ligra~\cite{shun13ppopp-ligra}, Galois~\cite{nguyen13sosp-galois} and GraphMat~\cite{sundaram15vldb-graphmat}.
Increasingly, many of these frameworks target a single multicore machine, because a single machine has the smallest communication overhead and memories have grown to the point where many graphs can fit on one server~\cite{shun13ppopp-ligra,mcsherry15hotos}.

\punt{
Given the interest in this field, it is natural to ask whether current
frameworks are close to hardware limits.
Perhaps surprisingly, we find that they are not: several optimizations can improve performance by 2--4$\times$ on common applications.
}

Given the effort in this field, it is natural to ask whether current frameworks are close to hardware limits. Perhaps surprisingly, we find that they are not. \Cagra employs novel optimizations to achieve up to 5$\times$ speedup over state-of-the-art shared memory graph frameworks.

The core problem is that graph applications have poor cache
utilization. They do very little computation per byte accessed, and a
large fraction of their memory requests are random.
Random accesses to a working set that does not fit in cache make the
entire cache hardware subsystem ineffective. Without effective use of the cache to mitigate the processor-DRAM gap,
CPUs are stalled on high-latency random accesses to DRAM.  Indeed, we
find that today's fastest frameworks spend 60--80\% of their cycles
stalled on memory access.

\punt{
\begin{figure}[t]
  \centering
	\includegraphics[width=\columnwidth]{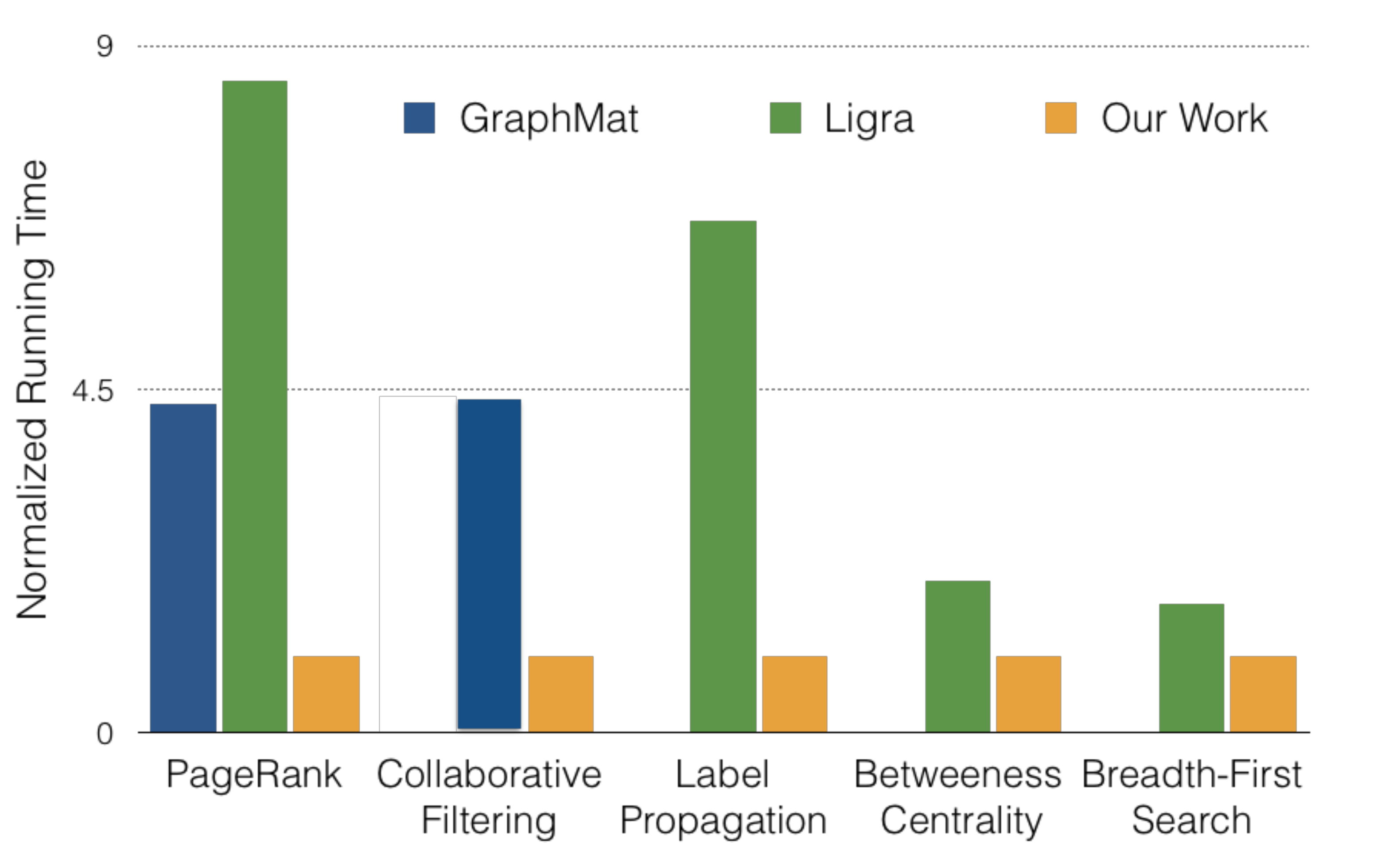}
	\caption{Running time with our techniques vs.~best published implementations in current frameworks on RMAT27.}
	\label{fig:speedup-allapps}
\end{figure}
}

The fastest in-memory frameworks, such as GraphMat and Ligra, do not optimize for caches aggressively. On the other hand, recent disk-based graph frameworks have applied techniques developed for the memory/disk boundary to the cache/DRAM boundary. However, we find that even the fastest of these frameworks, GridGraph~\cite{Zhu15ATC-GridGraph}, is 3$\times$ slower than  in-memory frameworks that do not optimize for caches (e.g., GraphMat). 

The main challenges in applying cache optimizations are achieving good multicore scalability and keeping the runtime overhead low enough to work with the smaller gap between cache and DRAM. For example, GridGraph~\cite{Zhu15ATC-GridGraph}, which uses a cache friendly 2D grid representation and a scatter-apply execution model, has trouble scaling efficiently beyond 4--6 cores. X-Stream~\cite{roy13sosp-xstream} partitions the graph into streaming partitions that fit in cache and processes each partition with a scatter-gather execution model. However, it incurs significant runtime overhead from the additional shuffle and gather phases. To make caches work for graph analytics, we need to carefully design multiple aspects of the system, such as partitioning the graph (2D Grid, Streaming Partitions or other schemes), choosing a data format (sorted compressed graph or unsorted edge list), exploiting parallelism (either within a single partition or parallelism across multiple partitions), utilizing the multi-level cache system (per core L1, L2 or shared LLC), and minimizing overhead. Many of these decisions are also interrelated, adding another layer of complexity to the problem.

In this paper, we present \Cagra, a framework that significantly speeds up graph applications on multicores using a new scalable and low-overhead technique to optimize cache usage.
The core of \Cagra is \emph{compressed sparse row (CSR) segmenting}, a novel partitioning and computation design that limits all random accesses in the system to the cache, and makes all DRAM accesses sequential. Unlike disk-based systems, \Segmenting uses a compressed 1D-segmented graph representation to improve the scalability of parallel in-cache processing and reduce overhead. \Segmenting first preprocesses the graph to divide the vertex data into cache-sized segments, and partitions the edges into subgraphs based on these segments. It then processes each subgraph in parallel, making one pass through all the edges while keeping the random accesses to vertex data in the cache. The intermediate updates from each segment are locally merged and stored using a buffer to eliminate random writes to DRAM. Finally, \Segmenting employs a novel low overhead parallel cache-aware merge algorithm to combine the updates from all the buffers within L1 cache. With \segmenting, \Cagra achieves up to 5$\times$ speedup over previous optimized graph frameworks, including cache-optimized techniques such as GridGraph and Hilbert Curve Ordering~\cite{mcsherry15hotos}. 

\punt{
abandons the 2D-paritioned graph scheme used by frameworks such as GridGraph~\cite{Zhu15ATC-GridGraph} and
}

\Segmenting can also be combined with \Reordering, a variant of degree based graph reordering technique, to further boost cache line utilization and keep frequently accessed vertices in fast cache. In many graph applications, each random access brings only one useful vertex data in each cache line. Taking advantage of the power-law degree distribution, we pack popular vertices together in memory to improve overall cache line utilization. \Cagra first applies \reordering to the entire graph and then performs \segmenting.

We evaluate \Cagra with various applications on large graphs with skewed degree distributions, as found in real world social, web and rating graphs. \Cagra achieves 5$\times$ speedup for PageRank, Label Propagation, 4$\times$ speedup for Collaborative Filtering, and 2$\times$ for Betweenness Centrality over the fastest previous frameworks. 

\punt {
We implement our techniques in a framework exposing an interface similar
to Ligra, and we demonstrate their contributions to significant
performance gains on a variety of representative applications compared to
the best published results. As shown in Figure~\ref{fig:speedup-allapps}, our optimizations provide
up to a 4.3$\times$ speedup for PageRank over Intel's GraphMat, 8.8$\times$
over Ligra, up to 6$\times$ speedup for Label Propagation, 4$\times$ speedup for Collaborative Filtering, and 2$\times$ for Betweenness Centrality.

We evaluate \Cagra with a number of graph application running on both real-world and synthetic graphs and demonstrate up to 5$\times$ performance gains over state-of-the-art in-memory and cache optimized frameworks and 3$\times$ over expert optimized hand-coded C++ baselines. We also see similar speedups over the recently proposed Hilbert curve ordering for graph data~\cite{mcsherry15hotos}.

}

\punt{ , as we discuss in Section~\ref{sec:eval-orderings}.\footnote{
In essence, although Hilbert order gives an effective single-threaded cache-oblivious algorithm, on multicores each core loads a different portion of the graph into cache. In contrast, our segmenting technique lets all cores share the \emph{same} working set in the cache.
}
}

\begin{figure}[t]
	\centering
		\includegraphics[width=0.8\columnwidth]{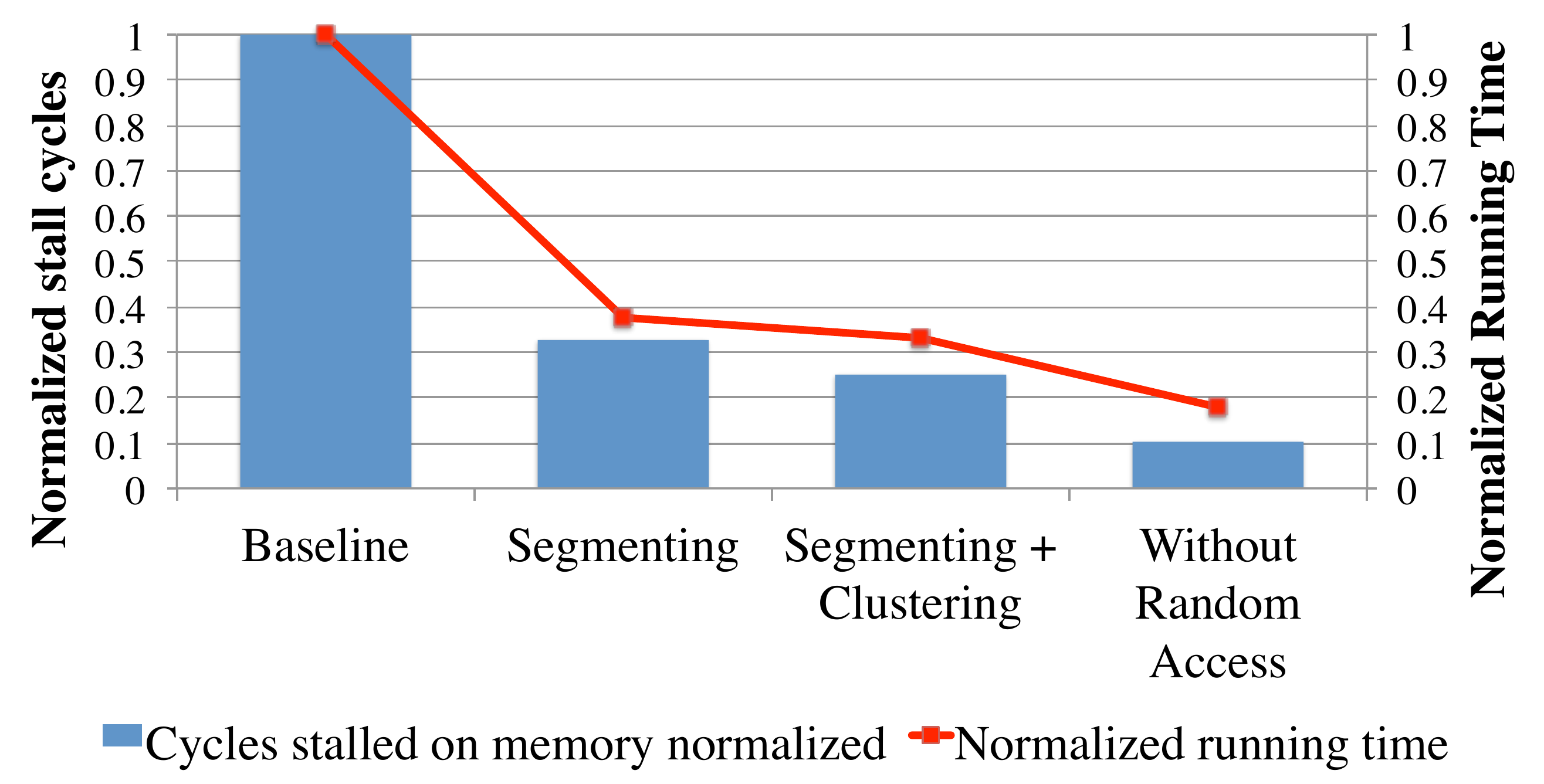}
		 \caption{Normalized running time and cycles stalled on memory 
                   for \Cagra's optimizations in PageRank on
                   the RMAT27 graph. The last bar is an incorrect program where
                   random accesses were removed to provide a lower bound.}
	    \label{fig:speedup-PageRank-rmat}
\end{figure}

Figure~\ref{fig:speedup-PageRank-rmat} explains our speedups further by showing how \segmenting and clustering reduce the cycles stalled on memory for PageRank. The last bar shows a modified
version of PageRank where all reads go to vertex 0 with \emph{no} random access to DRAM; \Cagra is within 2$\times$ of this lower bound.

\punt {
analyzes the speedup on PageRank. 
We see
Combining Segmenting with Clustering gives additional performance gains.
 (but the result is incorrect)
}

\punt {
Cache-aware segmenting makes better use of the cache hierarchy by grouping the vertices into \emph{segments} that are each loaded into the last-level cache once and then used in multiple random accesses.
That is, the random accesses that an algorithm makes each iteration are grouped by segment: we first load segment 1 into cache and perform accesses against it, then segment 2, and so on.
With this technique, most random accesses that would previously have hit DRAM now hit the last-level cache, and access to DRAM is sequential.
Although this technique adds some bookkeeping and merging cost, this cost is worth the reduction in random access.
}

In summary, our contributions are:
\begin{denseitemize}
\item We propose \Segmenting, which partitions graph data in a novel way and uses a cache-aware merge to process the intermediate updates. \Segmenting eliminates all random access to DRAM and makes all DRAM access sequential, achieving scalable cache-efficient graph processing.
\item We present \Cagra, a new framework that offers a programming interface to automatically apply \Segmenting to various graph applications. 
\item We evaluate \Cagra on several representative applications and demonstrate significant speedup over best published results. We also provide a detailed analysis of \Cagra's cache performance with stall cycles analysis and compare \Cagra with cache optimized GridGraph and Hilbert Curve Ordering techniques. 
\end{denseitemize}

%% file: understanding.tex
\section{Motivation}
\label{sec:understanding}

We will use PageRank listed in Algorithm~\ref{alg:PullPageRank} as a running example to motivate our optimizations for graph processing. PageRank iteratively updates the rank of each vertex based on the rank and degree of its neighbors. The performance characteristics of PageRank can generalize to a large number of graph applications.

\punt {

PageRank exemplifies many other graph algorithms, where a large portion of the
time is spent in random access to data, and there is little computation per byte.
For example, Breadth-First Search (BFS), Betweenness Centrality (BC), and
Single Source Shortest Paths (SSSP) have similar access patterns and similar
amounts of computation per byte accessed.\footnote{
Some of these algorithms differ from the PageRank one shown here in that they
track an \emph{active set} of vertices instead of updating every vertex on each
iteration.
We will evaluate two such algorithms, BFS and BC, in Section~\ref{sec:evaluation}.
}

}

\punt {
PageRank takes as input a graph $G$, a damping factor $d$, and a convergence
constant $t$.
It initializes the rank of each vertex to 1 and iteratively updates their ranks based on the ranks of their neighbors.
On each iteration, the algorithm updates the rank of vertex $i$ to
$ \mathrm{newRank}_i = (1-d) + d \sum_{j \in N(i)} \frac{\mathrm{rank_j}}{\mathrm{degree_j}}, $
where $N(i)$ is the list of vertices with edges to $i$. The algorithm stops when the ranks
change less than $t$ on an iteration.
}

\subsection{Graph and Vertex Data Representation}
\label{sec:understanding-data}

Graph frameworks typically store graph in Compressed Sparse Row (CSR)
format. Assuming the graph has V vertices and E edges, CSR format
would create a vertex array, \texttt{G.vertexArray}, of of length $
O(V)$ and an edge array, \texttt{G.edgeArray}, of size $ O(E)$. Vertex
Array stores the indices of the first neighbor of each vertex in the
Edge Array and use that to access the neighbor list of each
vertex. Application specific data is stored as separate arrays. In the
case of PageRank, vertex data is stored as arrays \texttt{newRank}, \texttt{rank} and \texttt{degree}
 of length $O(V)$.

\subsection{Memory Access Pattern}
The algorithm sequentially reads size $ O(E)$ data. By going over every vertex in order, the algorithm issues sequential read requests to \texttt{G.edgeArray} and sequential writes requests to \texttt{newRank}. The algorithm randomly reads $O(E)$ times from size $ O(V)$ vertex data, including \texttt{rank} and \texttt{degree}. These read requests are random because we cannot predict the values of \texttt{u}. 

This pattern of sequentially accessed edge data and randomly accessed
vertex data is common in representative graph applications. Collaborative Filtering needs to randomly read each vertex's latent factors, and Betweenness Centrality needs to randomly access the active frontier and number of paths through each vertex.

\punt {
There are several ways to implement PageRank on a multicore system.
In this paper, we will largely focus on a \emph{pull} algorithm, listed in
Algorithm~\ref{alg:PullPageRank}, in which threads scan through portions of
the \texttt{newRank} array sequentially and randomly access values in the
\texttt{rank} array from the previous iteration.
Another option used sometimes is a \emph{push} pattern, where threads scan
sequentially through the \texttt{rank} array and make random updates to
\texttt{newRank}.

In this paper, we will largely focus on \emph{pull} algorithms; an example implementation of PageRank is listed in
Algorithm~\ref{alg:PullPageRank}. \emph{Pull} algorithms transpose the graph and update the destination of each incoming edge. Another option used sometimes is a \emph{push} pattern that does not transpose the graph and update the destination of outgoing edges. We chose the pull-based version because it is more scalable in multicore systems without the overhead of atomic writes and has been adopted by frameworks like Galois and Ligra.

}

\algrenewcommand{\alglinenumber}[1]{\footnotesize#1}

\algdef{SE}[SUBALG]{Indent}{EndIndent}{}{\algorithmicend\ }%
\algtext*{Indent}
\algtext*{EndIndent}

\algblockdefx[pfor]{ParFor}{EndParFor}[1]
  {\textbf{parallel for}~#1~\textbf{do}}
  {\textbf{end parallel for}}

\begin{algorithm}[t]
\caption{PageRank}
\label{alg:PullPageRank}
\begin{algorithmic}[1]
\small
\Procedure{PageRank}{Graph $G$}
        \ParFor{v : G.vertexArray}
           \For{u : G.edgeArray[v]}
           \State
              G.newRank[v] +=
               \Indent
            \State 
                G.rank[u] / G.degree[u]
               \EndIndent
           \EndFor
        \EndParFor
\EndProcedure
\end{algorithmic}
\end{algorithm}

\punt {
First we note that the computation per byte accessed ratio is low because the algorithm performs one divide and add for two integers and two doubles. 
}

\subsection{Random Memory Access Bottleneck}
\label{sec:bottlenecks}

Graph applications have poor cache hit rate and are largely stalled on
memory accesses, since the working set of realistic
graphs is much larger than the last level cache (LLC) of current
machines.  For example, the Twitter
graph~\cite{kwak10www-twitter} has 41 million vertices and 1.5 billion
edges. The \texttt{rank} and \texttt{degree} arrays, which together
form the working set that is randomly accessed, are 656 MB (assuming
64-bit doubles) and are many times larger than the
30--55 MB LLC of current CPUs. Even though there is a
higher than expected hit rate due to the power-law degree distribution
and the community structures in the graph \cite{Beamer15IISWC}, we
still find the LLC miss rate for PageRank to be more than 45\%.

As a result of the high cache miss rate, our performance profile shows
graph applications are spending 60-80\% of their cycles stalled on
memory access. Random
memory access becomes the major bottleneck because random access to
DRAM is 6-8x more expensive than random access to LLC or sequential
accesses to DRAM. Sequential access to DRAM effectively uses all
memory bandwidth, and benefits from hardware prefetchers
to further reduce latency.

%% file: segmenting.tex
\section{\Segmenting}
\label{sec:segmenting}

\Segmenting\ improves cache utilization by working on one cache-sized
\emph{segment} of vertex data at a time. To make \segmenting\ work for
the cache/memory hierarchy, we have to keep graph processing scalable across all cores and runtime overhead low with carefully designed preprocessing, segment processing and cache-aware
merging. This technique can be applied to a wide class of computations
that aggregate values over the neighbors of each vertex in the graph.

\punt {
We first explain segmenting for PageRank, and then describe
how to apply it in the Ligra programming model.
}

\begin{figure}[t]
 \centering
 
 \begin{subfigure}{\columnwidth}
 \centering
 \includegraphics[width=0.8\columnwidth, keepaspectratio=true]{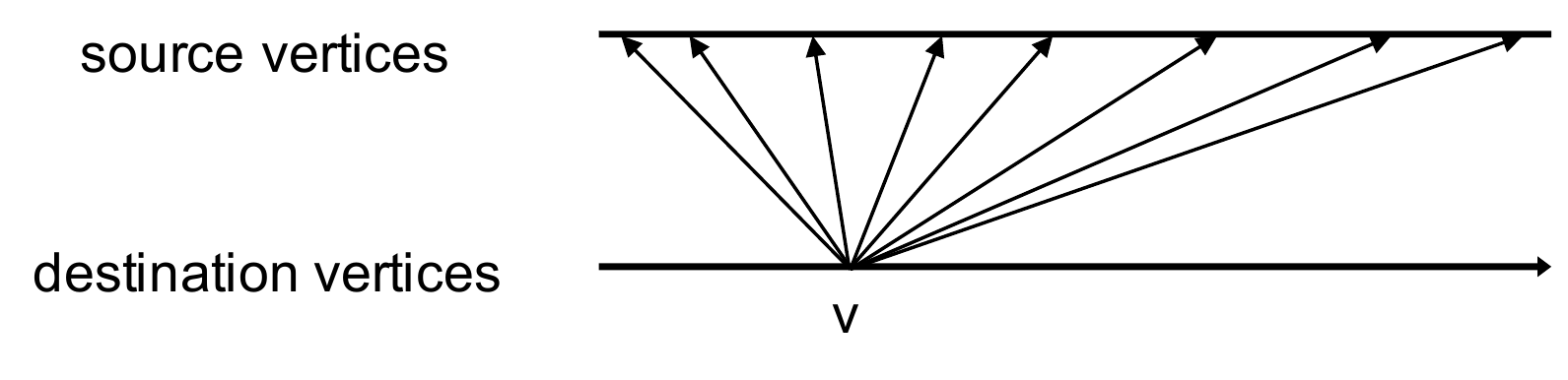}
 \caption{Without segmenting: we iterate through the destination
 vertices and read data from source vertices throughout the graph,
 causing random accesses to touch a large working set.}
 \label{fig:segmenting-before}
 \end{subfigure}
 
 \begin{subfigure}{\columnwidth}
 \centering
 \includegraphics[width=0.8\columnwidth, keepaspectratio=true]{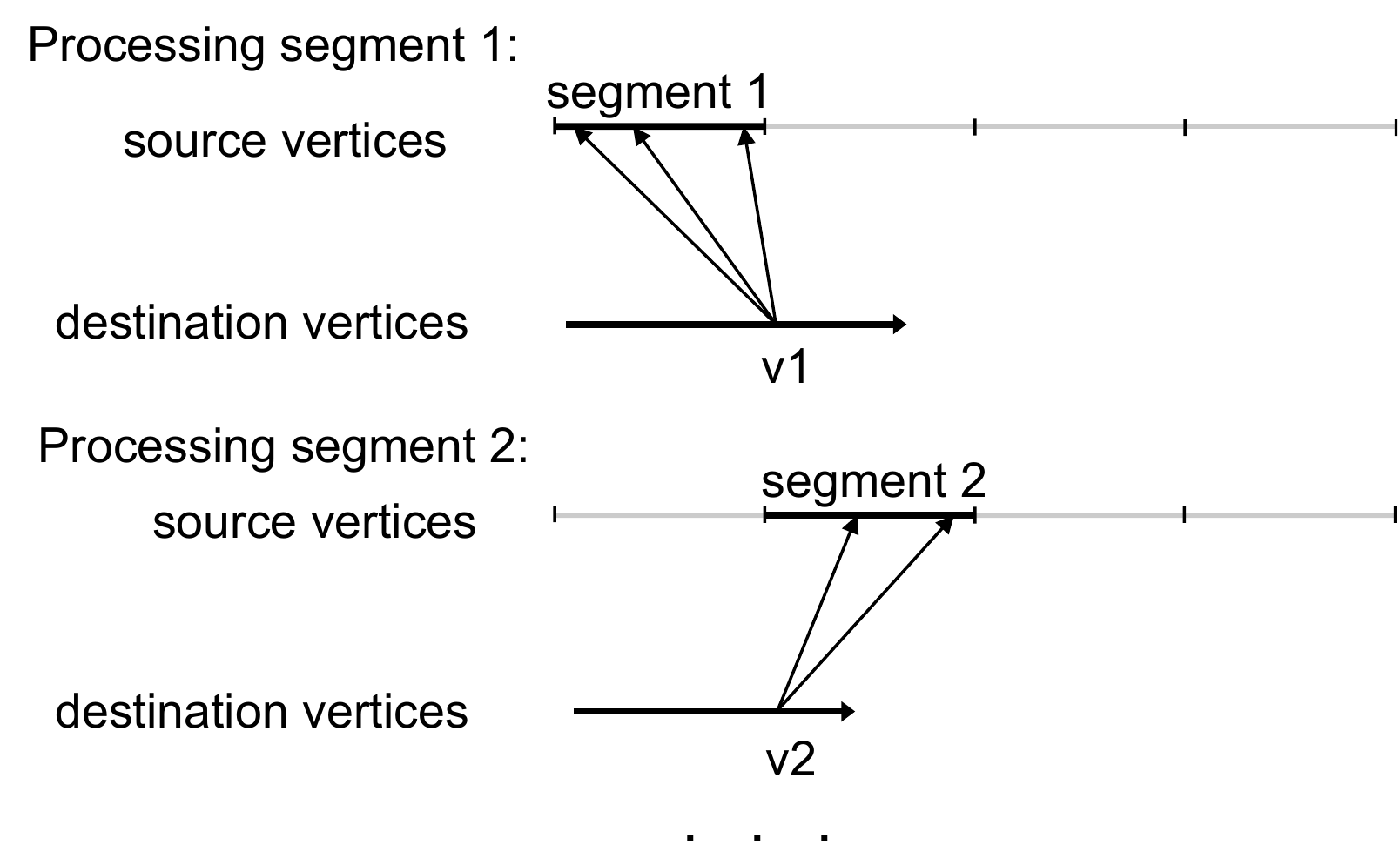}
 \caption{With segmenting, random accesses are now confined within a cache size segment.}
 \label{fig:segmenting-after}
 \end{subfigure}

 \caption{Processing each segment.}
 \label{fig:segmenting}
\end{figure}

\punt{
\begin{figure*}[t]
 \centering 
 \includegraphics[width=1.6\columnwidth]{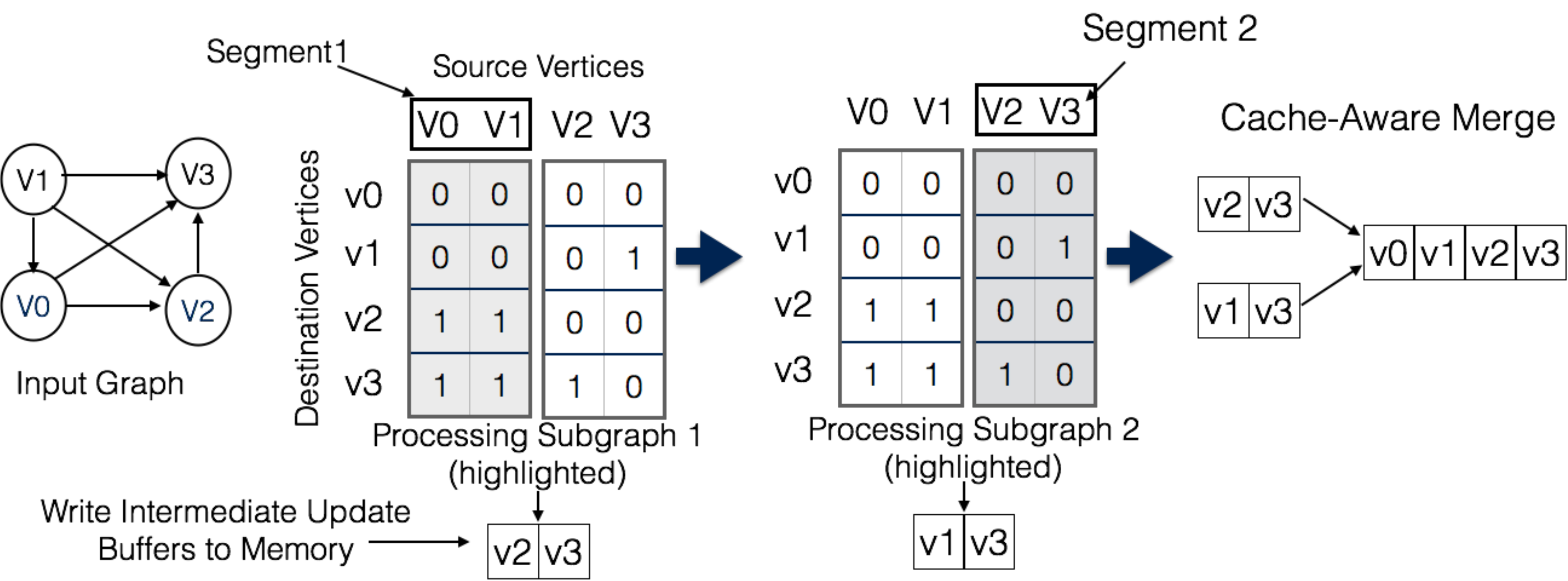}
 \caption{Example of applying \Segmenting. Two segments and their corresponding subgraphs are created from the input graph, which is represented as an adjacency matrix for demonstration (implemented as CSR). Subgraphs are processed in order and the intermediate update buffers are merged using parallel cache-aware merge. }
 \label{fig:segmenting_overall}
\end{figure*}
}

\begin{figure}[t]
 \centering
 \includegraphics[width=0.7\columnwidth]{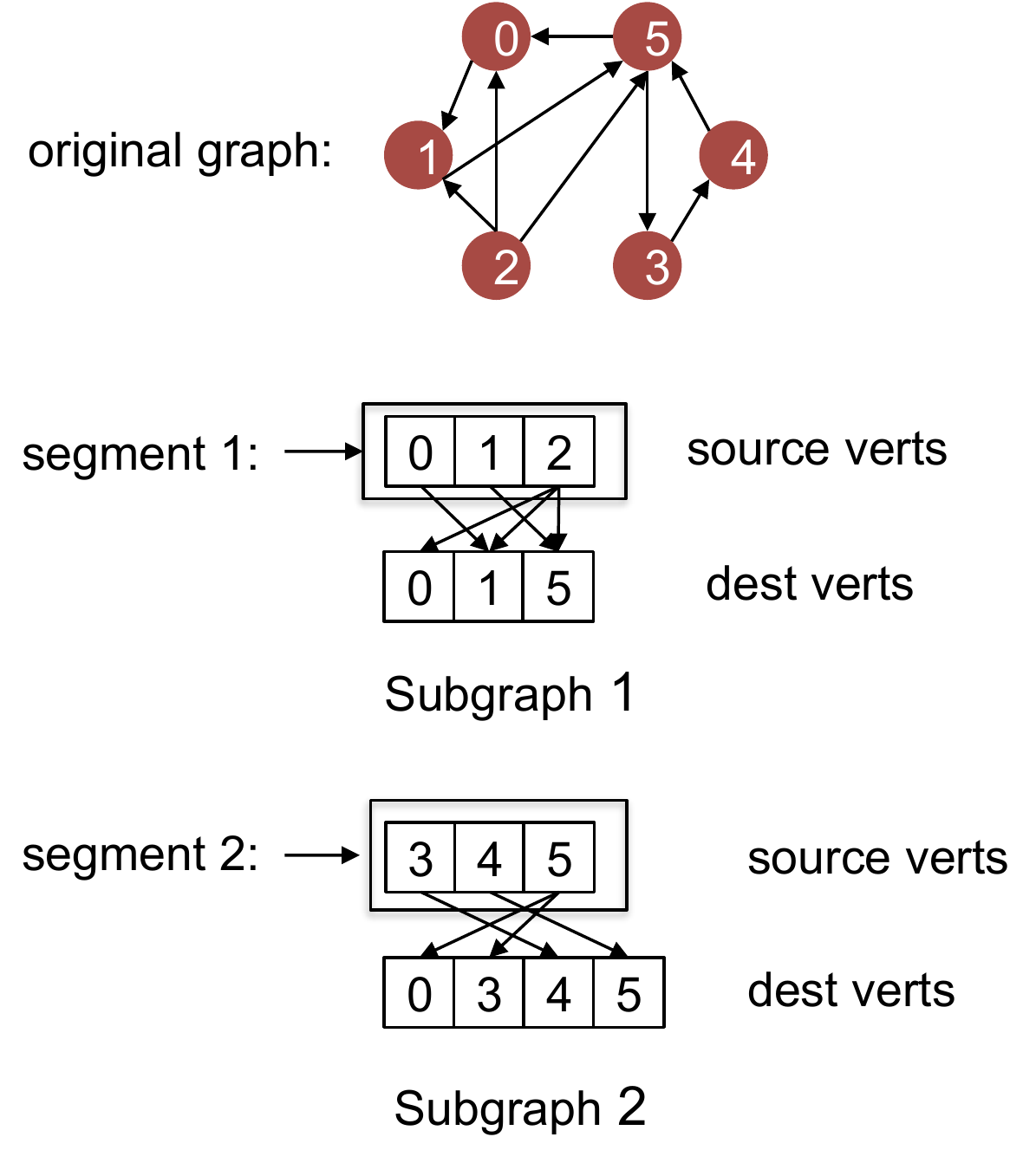}
 \caption{Example of subgraphs created in segmenting. We split the set of
 vertices, 0..5, into segments \{0,1,2\} and \{3,4,5\}, then build subgraphs with
 the destination vertices and out-edges for each segment.}
 \label{fig:segmenting-preproc}
\end{figure}

First, consider the PageRank algorithm in
Algorithm~\ref{alg:PullPageRank}. To compute the new ranks, each vertex randomly accesses a large array (the ranks of all vertices on the pervious iteration) to find its neighbors' rank. If this array does not fit in the CPU cache, many random accesses go to DRAM, shown in Figure~\ref{fig:segmenting}(a).

With segmenting, we partition the graph into subgraphs and make one pass through all the subgraphs. When processing each subgraph, we confine our random accesses to a cache-sized segment (Figure~\ref{fig:segmenting}(b)). Specifically, we first do a preprocessing step by dividing the previous iteration's rank array into $k$ segments that fit in the CPU's last-level cache. We then construct subgraphs by grouping together all the edges whose sources are in the same segment and construct a data structure for the destination vertices. \Segmenting\ processes one subgraph at a time. Within a subgraph, it iterates over all vertices in parallel and adds their contributions from the segment. Some destination vertices would potentially be duplicated across different subgraphs. As a result, we use parallel cache-aware merge to combine the contributions for each vertex from all segments that have edges to it, as shown in Figure~\ref{fig:merge}.

Segmenting can also potentially be applied to the push version of the algorithms. However, we focus on the InEdge processing (pull) version as it does not require atomic synchronizations, and takes most of the execution time~\cite{shun13ppopp-ligra}.

In summary, segmenting has the following benefits:
\begin{itemize}
  \item Improved cache utilization: It restricts all random reads and writes to cache, and makes all accesses to DRAM sequential. 
  \item Great scalability: Within each subgraph, threads can parallelize the execution across all vertices without using atomic operations for synchronization. Each subgraph is only partitioned on the source vertices, not on both the source and destination vertices, providing ample parallelism. The merge phase can be parallelized as well.
  \item Low overhead: Cache-aware merge only needs a small amount of extra sequential memory accesses and performs the merge in L1 cache in parallel. Processing the graph requires only one sequential pass through all the edges. 
  \item Easy to use: It can easily be applied to any algorithm that aggregates values across the graph with a clean API provided by \Cagra.
\end{itemize}

\punt {

In summary, segmenting has the following benefits:
\begin{itemize}
  \item It restricts all random accesses to cache, and makes all accesses
  to DRAM sequential. Even though extra work will be required for merging
  results, this work only needs sequential memory accesses and is less
  expensive than the random accesses saved.
  \item Within each segment, threads can parallelize work as before, and allfpara
  threads share all the data in the cache.
  \item It can easily be applied to any algorithm that aggregates values
  across the graph, as we will discuss by applying it inside Ligra.
\end{itemize}

}

We next describe and analyze the preprocessing, segment processing and cache-aware merge in more detail. 

\punt{, including preprocessing (Section~\ref{sec:segmenting-preproc}), computation within
a segment (Sections~\ref{sec:segmenting-inseg}), and our
cache-aware merge algorithm (Section~\ref{sec:segmenting-merge}).

We then discuss tradeoffs in choosing the size of segments
and the benefits of partitioning
after \reordering (Section~\ref{sec:segmenting-params}).
Finally, we discuss how to apply segmenting to other graph computations through
an extension of the Ligra API (Section~\ref{sec:segmenting-api}).}

\subsection{Preprocessing}
\label{sec:segmenting-preproc}

The preprocessing algorithm is shown in Algorithm~\ref{algo:preprocessing} and Figure~\ref{fig:segmenting-preproc} shows an example dividing a graph into
two subgraphs based on the segments. The first step is to construct the subgraphs based on the segments. We first divide the vertices into segments such that the data for each segment fits in the cache. For each segment $S$, we construct a new subgraph consisting of edges whose sources are in the segment. To do this, we compute the segmentID(subgraphID) of each inEdge by dividing the sourceID of the inEdge by the number of vertices in each segment $N$ and then add the edge to the subgraph. The edges in each subgraph are sorted by their destinations ($sortByDestination$). This step takes no time since the original graph in CSR is already sorted by destination. Then, a CSR representation will be constructed for each subgraph. The algorithm also creates an array, $intermBuf$, to hold the intermediate result for each destination vertex $v$. Additionally, we create an index mapping, $idxMap$, to map local index of destination vertices in the subgraph to their global index in the original graph. Finally, we create an index of blocks that stores block starts and ends used in cache-aware merge.   Note that this preprocessing phase can be done in parallel by building each subgraph separately from the original CSR.

\begin{algorithm}[t]
\small
\caption{Preprocessing}
\label{algo:preprocessing}
\begin{algorithmic}
\State \textbf{Input: } Number of vertices per segment N, Graph G
	\For{$v : G.vertices$}
		\For{$inEdge : G.inEdges(v)$}
         	\State $segmentID\gets inEdge.src / N$
         	\State $subgraphs[segmentID].addInEdge(v,inEdge.src)$
         \EndFor
	\EndFor
	
	\For{$subgraph : subgraphs$}
		\State $subgraph.sortByDestination()$
		\State $subgraph.constructIdxMap()$
		\State $subgraph.constructBlockIndices()$
		\State $subgraph.constructIntermBuf()$
	\EndFor
\end{algorithmic}
\end{algorithm}

\punt{
This process works as follows:
\begin{enumerate}
  \item Divide the vertices into segments such as the data for each segment
  fits in the cache. (Section~\ref{sec:segmenting-params} describes tradeoffs
  in the segment size, e.g., which level of the cache to use.)
  \item For each segment $S$, determine the vertices and edges that are adjacent
  to those in $S$ (i.e., edges with sources in $S$).  A new CSR will be constructed for the vertices in the segment and their adjacent edges. We also create an array
  to hold the intermediate result for each adjacent (destination) vertex $v$. 
  \item Create an index vector with the index of each adjacent (destination) vertex in the
  original graph, which will be used to combine intermediate results in the
  merge phase.
\end{enumerate}
}

\punt{
In our implementation we find that segmenting only takes time proportional to a few PageRank iterations, shown in Section~\ref{sec:preprocessingtime}.}

\subsection{Parallel Segment Processing}
\label{sec:segmenting-inseg}

After the preprocessing is done, the system processes each subgraph in turn, as shown in Figure~\ref{fig:segmenting}.
Within each subgraph, we parallelize the computation across different vertices. We made three key design choices in segment processing, shown in Algorithm~\ref{algo:segment-processing}. 

First we exploit parallelism within a single large segment that fits in LLC, instead of across multiple smaller segments. This way, all the threads in our approach share the \emph{same} working set, i.e., the source vertex data in this segment, which is read-only. Thus, adding more threads does not create cache contention. We also experimented with parallelizing the processing of multiple smaller segments. Each smaller segment's working set fits in L2 cache, instead of LLC, for even lower random access latency. However, we found that the merging overhead that comes with a large number of smaller segments becomes a significant performance bottleneck. 

Next, we divide the graph based on only source vertices, and not on both source the destination vertices (2D partitioning), to achieve good scalability. Many other frameworks, such as GridGraph, use 2D partitioning to make sure reads and writes happen in cache. However, this approach can create a large number of subgraphs with a small number of edges, resulting in poor scalability when processing each subgraph in parallel. Instead, we limit only the reads to be in cache, allowing writes to DRAM, as long as they are sequential.

Finally, we parallelize across different vertices, but not within the same vertex. This parallelization approach takes advantage of the CSR format of each subgraph to generate large degree of parallelism without using atomics for synchronization, since the updates to each vertex in the subgraph are locally merged by the same worker thread.

\punt{
By parallelizing work within a segment, we only need to create a relatively small number of segments, where each segment fits in the last level cache (LLC) and contains a large number of edges. This way, we keep the preprocessing time low, generate good parallelism and avoid high merging cost. Additionally, all the threads share the \emph{same} working set, i.e., the source vertex data in this segment, which is read-only. Thus, adding more threads does not create cache contention.
}

\algblockdefx[pfor]{ParFor}{EndParFor}[1]
  {\textbf{parallel for}~#1~\textbf{do}}
  {\textbf{end parallel for}}

\begin{algorithm}[t]
\small
\caption{Parallel Segment Processing}
\label{algo:segment-processing}
\begin{algorithmic}
	\For{$subgraph : subgraphs$}
		\ParFor{$v : subgraph.Vertices$}
         	\For{$inEdge: subgraph.inEdges(v)$}
         		\State \textbf{Process} $inEdge$
         	\EndFor
         \EndParFor
    \EndFor
\end{algorithmic}
\end{algorithm}

\begin{figure}[t]
 \centering
 \includegraphics[width=0.7\columnwidth]{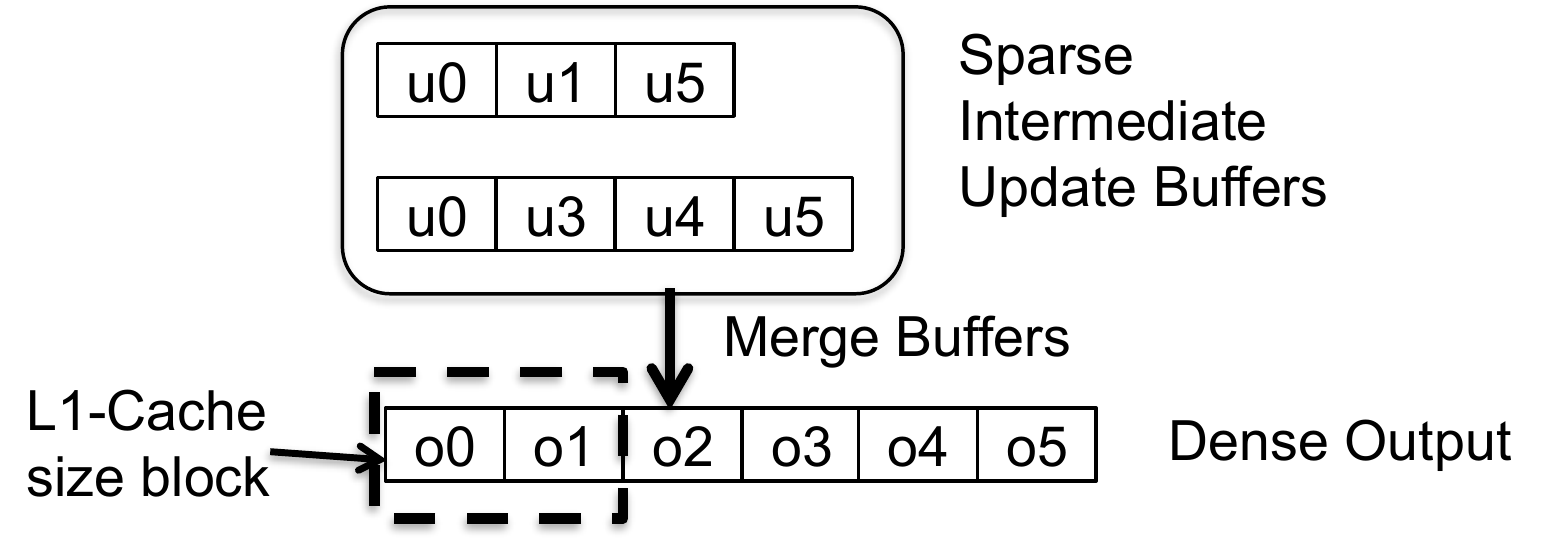}
 \caption{Cache-aware merge}
 \label{fig:merge}
\end{figure}

\subsection{Cache-Aware Merge}
\label{sec:segmenting-merge}

Once the per-segment passes are done, \Cagra fills up the intermediate
buffers for each subgraph with the updates for vertices. These buffers are sparse, holding data only for the
destination vertices in each segmented subgraph. For example, in
Figure~\ref{fig:segmenting-preproc}, segmented subgraph 1 will produce a
buffer with updates for vertices 0,1,5 and the second buffer for subgraph 2 will
produce a buffer with updates for vertices 0,3,4,5, as shown in Figure~\ref{fig:merge}.

To combine these intermediate update buffers into one dense output vector, we use a cache-aware merge algorithm (Algorithm~\ref{algo:merge}). The algorithm accesses the intermediate buffers sequentially, requiring
no branches, and runs in parallel.
We divide the range of vertex IDs into L1-cache-sized \emph{blocks}.
Then, for each block, a worker thread reads the updates for
that range of vertex IDs from the sparse buffers of intermediate
results, and updates a dense vector \emph{output} for the final output using the local to global index mapping \emph{idxMap}.
Helper data structures \emph{blockStarts} and \emph{blockEnds} hold the start and end local index of each output block's vertex IDs in each of the per-subgraph buffers, ensuring the random access is confined in L1 cache. Different blocks can be processed in parallel by different threads,
and we use a work-stealing load balancing scheme to divide them
across processors.\footnote{One benefit of this approach is that
each thread usually processes multiple consecutive blocks, further
increasing the range of sequential access for both reads and writes.}

\algrenewcommand{\alglinenumber}[1]{\footnotesize#1}

\algdef{SE}[SUBALG]{Indent}{EndIndent}{}{\algorithmicend\ }%
\algtext*{Indent}
\algtext*{EndIndent}

\algblockdefx[pfor]{ParFor}{EndParFor}[1]
  {\textbf{parallel for}~#1~\textbf{do}}
  {\textbf{end parallel for}}

\begin{algorithm}[t]
\small
\caption{Cache-Aware Merge}
\label{algo:merge}
\begin{algorithmic}
	\ParFor{$block : blocks$}
		\For{$subgraph : G.subgraphs$}
         	\State $blockStart\gets subgraph.blockStarts[block]$
         	\State $blockEnd\gets subgraph.blockEnds[block]$
         	\State $intermBuf\gets subgraph.intermBuf$
         	\For{$localIdx$ \textbf{from} $blockStart$ \textbf{to} $blockEnd$}
         		\State $globalIdx\gets subgraph.idxMap[localIdx]$
         		\State $localUpdate = intermBuf[localIdx]$
         		\State $\textbf{merge}(output[globalIdx], localUpdate)$
         	\EndFor
         \EndFor
    \EndParFor
\State $\textbf{return}$ $output$
\end{algorithmic}
\end{algorithm}

\begin{figure}[t]
 \centering
 \includegraphics[width=0.8\columnwidth]
{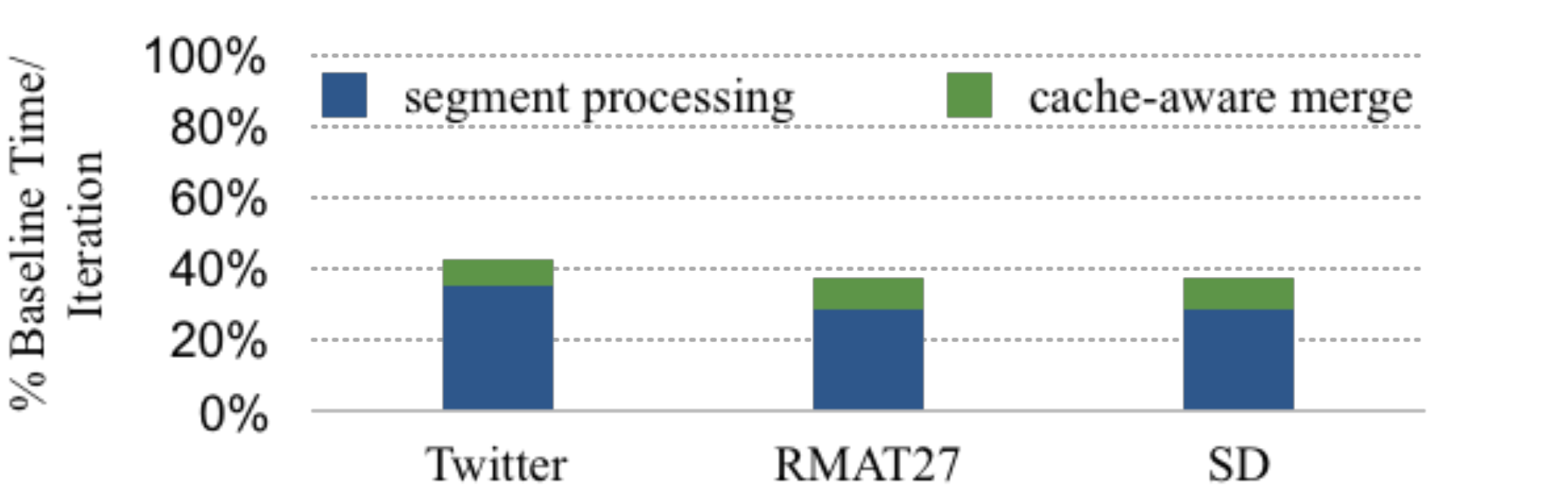}

 \caption{Comparison of segment computation vs merge costs.
   Runtime\% normalized to
   an optimized PageRank baseline without segmenting on 24 cores.}
 \label{fig:seg-breakdown}
\end{figure}

With the cache-aware merge algorithm, merging adds only a small runtime overhead. 
Figure~\ref{fig:seg-breakdown} shows the percentage of time
on segment processing and merge using 48 hyperthreads for
PageRank, normalized to a baseline with all our optimizations other
than segmenting.

\punt{
Other overhead includes all other time within each
iteration other than edge processing, e.g. the per-vertex division
to compute contributions.
}

\subsection{Segment Size Selection}
\label{sec:segmenting-params}

A final consideration in using segmenting is how large to make the segments.
In general, there is a tradeoff with segment size.
Smaller segments will fit into a lower-level cache (e.g., L1 or L2),
reducing random access latency.
However, smaller segments will also result in more merges for the same
destination vertex, because the source vertices pointing to it will be
in multiple segments.
Across the applications we evaluated, we found that sizing the segments
to fit in last level (L3) cache provided the best tradeoff.

\begin{figure}[t]
 \centering
 \includegraphics[width=0.8\columnwidth]
{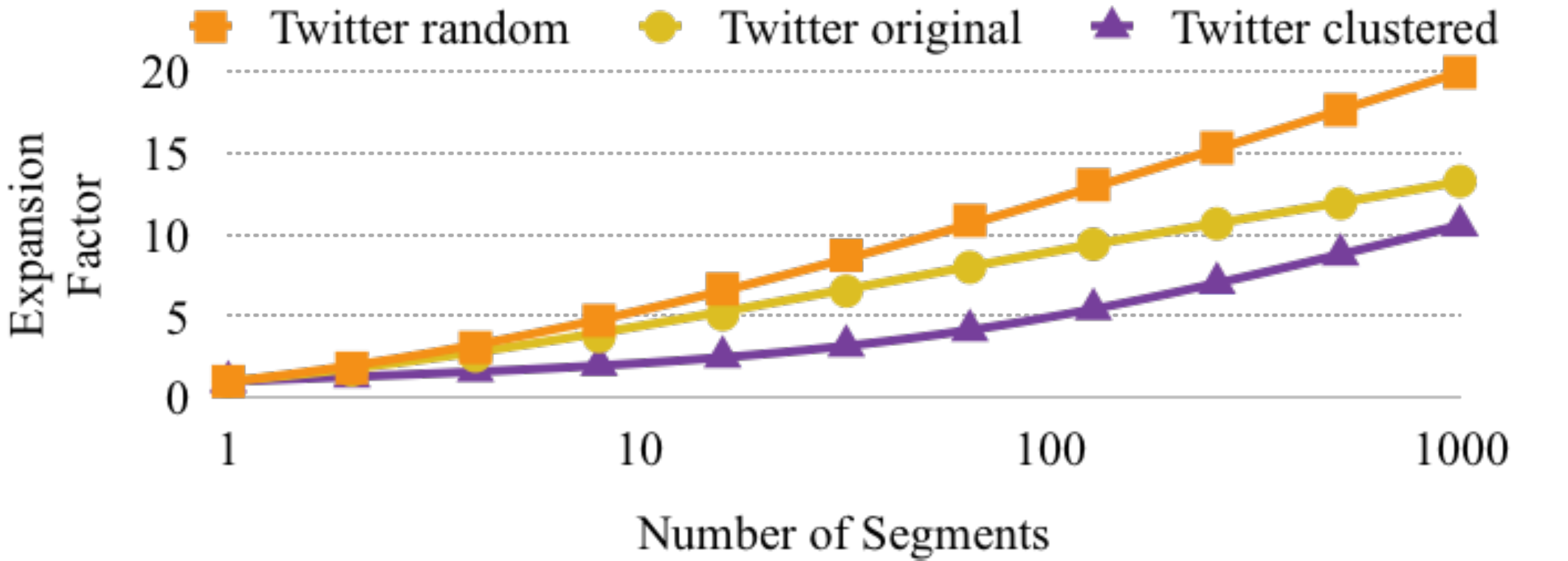}
 \caption{Expansion factors of segmenting Twitter graph in
   original and random order, frequency based clustered order.
}
 \label{fig:qfactor}
\end{figure}

To further understand the impact of segment size, we define a metric
called the \emph{expansion factor}.
Let $s$ be the number of vertices in each segment, and
$s_\mathrm{adj}$ be the average number of vertices adjacent to
each segment, that is, with edges from the segment to them.
Then we define $q=s_\mathrm{adj} / s$ as the expansion
factor.
The expansion factor describes how many segments, on average,
contribute data to each vertex, and hence how many merge operations
happen for each vertex.

Figure~\ref{fig:qfactor} shows the expansion factors as a function of
number of segments for several graphs while varying number of
vertices, average degrees, and vertex order.  For PageRank
calculations where we need 8-byte data per vertex, a 30 MB LLC cache
can fit 4M vertices.  For these workloads the expansion factors are less
than 5, which is much less than the number of
segments or the average degree (24 for the Twitter graph) which are the upper bounds
of $q$.  Randomly permuting vertices results in a much worse expansion factor.

\subsection{Analysis of Memory Access Costs}
\label{sec:segmenting-analysis}
We analyze the memory access efficiency of \segmenting by analyzing the total traffic between LLC and DRAM and the total number of cache misses in LLC. 

\textbf{Traffic between LLC and DRAM: } Assume we have $k$ segments, and the expansion factor detailed in the last section is $q$. When processing each segment, we only need to read in $V/k$ source vertex data and write $qV/k$ intermediate updates buffer to memory. \Cagra makes only one pass through all the edges. As a result, it incurs a total of $E + qV + V$ traffic during segment processing phase. In cache-aware merge, we first read back all the intermediate buffers ($qV$) to perform the merge and write back $V$ final values, incurring $qV + V$ traffic. Summing up both phases, we get the total DRAM traffic:
 \centerline{ $E + 2qV + V$ }

 We compare \Cagra's sequential and random memory traffic with other cache optimized frameworks, including GridGraph and X-Stream, in Section~\ref{sec:related}.

%% file: programming.tex
\section{Programming Abstraction}

\begin{algorithm}[t]
\small
\caption{PageRank in Cagra}
\label{algo:pr_ex}
\begin{algorithmic}

\State typedef double vertexDataType
\State $contrib\gets $ \{1/outDegree[v], ...\} 
\State $newRank\gets $ \{0.0, ...\} \\

\Procedure{EdgeUpdate}{$bufVal,srcVal,dstVal$}
        \State $bufVal += srcVal$
        \State \textbf{return true}
\EndProcedure \\

\Procedure{Merge}{$newDstVal,bufVal$}
        \State $newDstVal += bufVal$
\EndProcedure \\

\Procedure{VertexUpdate}{$v$}
		\State $newRank[v]\gets 0.15+0.85*newRank[v]$
		\State $newRank[v]\gets newRank[v]/outDegree[v]$
		\State $contrib[v]\gets 0.0$
        \State \textbf{ return true }
\EndProcedure \\

\Procedure{PageRank}{$G, maxIter$}
		\State $iter\gets 0$
		\State $A\gets V$
		\While {$iter \not=maxIter$}
			\State $A\gets EdgeMap(G, A, EdgeUpdate, EdgeMerge)$
			\State $A\gets VertexMap(G, A, VertexUpdate)$
			\State $Swap(contrib, newRank)$
			\State $iter\gets iter+1$
		\EndWhile
\EndProcedure 
\end{algorithmic}
\end{algorithm}

\Cagra extends the EdgeMap and VertexMap API from Ligra to support automatic cache optimizations. Cagra processes a directed graph $G=(V,E)$, where V is the vertex set and E is the edge set. Undirected graph is represented with edges in both directions. Algoirithm~\ref{algo:pr_ex} demonstrates the pseudocode of PageRank implemented in Cagra.

\textbf{EdgeMap (G, ActiveFrontier, EdgeUpdate, Merge)}
Similar to the original Ligra API, {\small\tt EdgeMap} traverses the edge set E, and applies an {\small\tt EdgeUpdate} function to the edge if the source node is in the ActiveFrontier. \Cagra requires users to provide a {\small\tt Merge} function to perform cache-aware merge, as shown in Algorithm~\ref{algo:merge}. Furthermore, the update function works with data values of source and destination vertices, instead of their indices as in Ligra, since the indices in the original graph's CSR are different from those in the CSR of the current segmented graph. The users direct all writes to the segmentVal. 

\punt{We also provide access to OldDstVal for applications that need to read the old value of the destination vertices to perform the updates, such as Collaborative Filtering.}

\textbf{VertexMap (G, VertexSubset, VertexUpdate)} 
VertexMap applies the user defined VertexUpdate function to every vertex in the VertexSubset.

%% file: reorder.tex
\section{\Reordering}
\label{sec:reorder}



\Segmenting can also be combined with \Reordering, an out-degree based graph reordering technique, to further boost cache line utilization and keep frequently accessed vertices in fast cache. \Reordering reorganizes the physical layout of the vertex data structures to improve cache utilization. It reduces overall cycles stalled on memory by serving more random requests in fast storage.

\punt{ reorganizes the physical layout of the vertex data structures to improve cache utilization. It reduces overall cycles stalled on memory by serving more random requests in fast storage. This technique is often effective when used alone, but can also further boost \segmenting when combined together. 
}

\textbf{Motivation:} We make three key observations on graph access patterns and motivate \reordering. First, each random read in graph applications often only uses a small portion of the cache line. For PageRank, the size of the vertex data is 8 bytes for a rank represented as a double, using only 1/8 of a common 64 byte cache line. Second, certain vertices are much more likely to be accessed than others in power law distributed graphs, where a small number of vertices have a large number of edges attached to them~\cite{kwak10www-twitter}. A third observation is that the original ordering of vertices in real world graphs often exhibit some locality. Vertices that are referenced together are sometimes placed close to each other due to existing communities.

\textbf{Design and Implementation:} We designed \reordering to group together the vertices that are frequently referenced, while preserving the natural ordering in the real world graphs. We use out-degrees to select the frequently accessed vertices because many graph algorithms use only pull based implementations, or spend a significant portion of the execution time in the pull phase. To preserve the original ordering in real world graphs, we cluster together only vertices with out-degree above the average degree of nodes. This thresholding allows us to keep some of the locality in the original ordering, yet still offering a clustering of high-out-degree vertices that maximizes the effectiveness of L1, L2, and L3 caches.

\punt{
We use out-degrees to select the frequently accessed vertices because many state-of-the-art implementations of graph algorithms use only pull based implementations, or spend a significant portion of the execution time in the pull phase, such as BFS and BC with direction optimization~\cite{Beamer-2012}. Furthermore, if we optimize for the push phases by using in-degree of vertices, we incur significant penalty from false sharing when a large number of write requests are directed to the same cache line. 

To preserve the original ordering in real world graphs, we cluster together only vertices with out-degree above a threshold, leaving the ordering of other vertices intact. A good threshold is the average degree of nodes. This thresholding allows us to keep some of the locality in the original ordering, yet still offering a cache-oblivious clustering of high-out-degree vertices that maximizes the effectiveness of L1, L2, and L3 caches. \punt{ More analysis is shown in Section~\ref{sec:eval-orderings}.}
}

We use a parallel stable sort based on vertices' out-degree/threshold to cluster together frequently referenced vertices. Next, we create a mapping from old vertex index to the newly sorted vertex index and use the mapping to update the vertex index in the \texttt{G.edgeArray}. Load balance is critical to achieving high performance with \reordering. The thread responsible for the part of the vertex array containing high out-degree vertices may perform much more work than other threads. We implemented a work-estimating load balancing scheme that partitions the vertex array based on the number of edges within each task, which reflects how many random reads it will make to the \texttt{rank} array. The task then processes its range of vertices if the cost is sufficiently small, or divides into two sub-tasks otherwise.

\punt{
Many in-memory graph frameworks, including Ligra, see significant slow down with \reordering, using the default vertex based load balance scheme.}

\textbf{Combining Clustering and \Segmenting:} Segmenting works well with \reordering. \Cagra first applies \reordering on the entire graph and then proceeds to apply \segmenting. The first advantage is that \Cagra can now make better use of faster higher level caches. Additionally, the clustered graph requires less extra sequential memory overhead. The expansion factors shown in Figure~\ref{fig:qfactor} for the clustered Twitter graph is over 2$\times$ smaller than the original graph, reducing sequential memory traffic overhead.

%% file: evaluation.tex
\section{Evaluation}
\label{sec:evaluation}

We demonstrate up to 5$\times$ speedup for various graph applications over best published results from state-of-the-art graph processing frameworks, and 3$\times$ speedup over previously expert hand optimized C++ implementations. We provide a detailed analysis on cycles stalled on memory to show that \Cagra's improved cache efficiency. Finally, we show \Cagra's good scalability and low runtime overhead through comparisons with other cache optimized frameworks and techniques, including GridGraph and Hilbert Curve Ordering.  
\subsection{Experimental Setup}

We conducted our experiments on a dual socket system with Intel Xeon
E5-2695 v2 CPUs 12 cores for a total of 24 cores and 48
hyperthreads, with 30 MB last level cache in each socket. The system
has 128GB of DDR3-1600 memory. The machine runs with Transparent Huge Pages (THP) enabled. 

\punt{ For parallel execution, we used Intel's Cilk Plus compiler and runtime system. All code is optimized at highest optimization with \texttt{-O3 -ipo} flags. We used the best configurations noted by Ligra and GraphMat release. }

\punt{

Ubuntu 14.04, with
Linux kernel 3.13, 

from Intel C++ Composer XE 2015 v15.0.3.

We used Ligra released as of Sep 11th 2015 and
GraphMat 1.0 for performance comparisons. 
 While theoretical peak memory
system bandwidth is 102GB/s, practically achievable sequential read bandwidth from each
socket is 47GB/s, and when using NUMA interleaving the peak total
bandwidth is 79GB/s.

We collected cycles stalled on memory for the applications using \texttt{perf}.

}


\textbf{Data Sets}: We used the social networks, including LiveJournal~\cite{davis11acm-florida-sparse} and Twitter~\cite{kwak10www-twitter}, the SD web graph from 2012 common crawl~\cite{sd-graph}, and the RMAT graphs. We also synthesized an expanded version of the Netflix dataset.Table~\ref{table:datasets} summarizes the datasets that we use.

\begin{table}[t]
\center
\tabcolsep 5pt
\footnotesize
\begin{tabular}{l|r|r}
\hline
Dataset & Number of Vertices &  Number of Edges \\ \hline 
LiveJournal~\cite{davis11acm-florida-sparse} & 5 M & 69 M  \\ \hline
Twitter~\cite{kwak10www-twitter} & 41 M & 1469 M \\ \hline
\textit{RMAT 25}~\cite{ChakrabartiSDM-rmat} & 34 M & 671 M \\ \hline
\textit{RMAT 27}~\cite{ChakrabartiSDM-rmat} & 134 M & 2147 M \\ \hline
\textit{SD}~\cite{sd-graph} & 101 M & 2043 M \\ \hline
{Netflix}~\cite{Bennett07thenetflix} & 0.5 M & 198 M  \\ \hline
\textit{Netflix2x}~\cite{LiNetflix} & 1 M & 792 M \\ \hline 
\textit{Netflix4x}~\cite{LiNetflix} & 2 M & 1585 M \\ \hline
\end{tabular}
\caption{Real world and \textit{synthetic} graph input datasets}
\label{table:datasets}
\end{table}



\textbf{Applications}: We choose a representative set of applications from domains such as machine learning, graph traversals and graph analytics. PageRank, Label Propagation and Collaborative Filtering are dominated by unpredictable vertex data accesses. The algorithms do not require any vertices' activeness checking. Additionally, PageRank, Label Propagation and Collaborative Filtering take a number of iterations to complete, justifying the preprocessing time. Betweenness Centrality (BC) represents the applications that involve vertices' activeness checking and making unpredictable access to vertices' data, such as single source shortest path (SSSP). Betweenness Centrality also takes a large number of iterations, making a case for additional preprocessing time. 
\punt{
Breadth First Search represents a unique class of applications that only need to do activeness checking on the vertices without accessing vertices' data. This application has the smallest working set in graph applications. 
}

\begin{table}[t]
\center
\tabcolsep 5pt
\footnotesize
\begin{tabular}{p{0.9cm}|p{1.1cm}|p{1.0cm}|p{1.1cm}|p{0.9cm}|p{1.1cm}}
\hline
Dataset & \Cagra & \Baseline & GraphMat & Ligra & GridGraph\\ \hline 
Live Journal  & 0.017s (1.00$\times$) & 0.031s (1.79$\times$) & 0.028s (1.66$\times$) &
0.076s (4.45$\times$) & 0.195 (11.5$\times$)\\ \hline
Twitter &  0.29s  (1.00$\times$) & 0.79s (2.72$\times$) & 1.20s  (4.13$\times$) & 2.57s (8.86$\times$) & 2.58 (8.90$\times$) \\ \hline
RMAT 25 &  0.15s (1.00$\times$) & 0.33s (2.20$\times$) & 0.5s (3.33$\times$) & 1.28s (8.53$\times$) & 1.65 (11.0$\times$) \\ \hline
RMAT 27 &  0.58s (1.00$\times$) & 1.63s (2.80$\times$) & 2.50s (4.30$\times$) & 4.96s (8.53$\times$) & 6.5 (11.20$\times$) \\ \hline
SD & 0.43 (1.00$\times$) & 1.33 (2.62$\times$) & 2.23 (5.18$\times$) & 3.48 (8.10$\times$)  & 3.9 (9.07$\times$) \\ \hline
\end{tabular}
\caption{PageRank runtime per iteration comparisons with other frameworks
  and slowdown relative to \Cagra }
\label{table:PageRankComparisons}
\end{table}

\begin{table}[t]
\center
\tabcolsep 5pt
\footnotesize
\begin{tabular}{l|r|r|r}
\hline
Dataset & \Cagra & \Baseline & GraphMat  \\ \hline 
Netflix &  0.20s  (1$\times$) & 0.32s (1.56$\times$) & 0.5s  (2.50$\times$) \\ \hline
Netflix2x &  0.81s (1$\times$) & 1.63s (2.01$\times$) & 2.16s (2.67$\times$) \\ \hline
Netflix4x &  1.61s (1$\times$) & 3.78s (2.80$\times$) & 7s (4.35$\times$) \\ \hline
\end{tabular}
\caption{Collaborative Filtering runtime per iteration comparisons with GraphMat
  and slowdown relative to \Cagra }
\label{table:CFComparisons}
\end{table}

\begin{table}[t]
\center
\tabcolsep 5pt
\footnotesize
\begin{tabular}{l|r|r|r}
\hline
Dataset & \Cagra & \Baseline & Ligra \\ \hline 
Live Journal  & 0.02s (1$\times$) & 0.01s (0.68$\times$) & 0.03s (1.51$\times$) \\ \hline
Twitter &  0.27s  (1$\times$) & 0.51s (1.73$\times$) & 1.16s  (3.57$\times$) \\ \hline
RMAT 25 &  0.14s (1$\times$) & 0.33s (2.20$\times$) & 0.5s (3.33$\times$) \\ \hline
RMAT 27 &  0.52s (1$\times$) & 1.17s (2.25$\times$) & 2.90s (5.58$\times$) \\ \hline
SD & 0.34 (1$\times$) & 1.05 (3.09$\times$) & 2.28 (6.71$\times$) \\ \hline
\end{tabular}
\caption{Label Propagation runtime per iteration comparisons with other frameworks
  and slowdown relative to \Cagra }
\label{table:LPComparisons}
\end{table}

\begin{table}[!h]
\center
\tabcolsep 5pt
\footnotesize
\begin{tabular}{p{1.5cm}|r|r}
\hline
Dataset & \Cagra & Ligra \\ \hline 
LiveJournal  & 1.2s (1$\times$) & 1.2s (1.00$\times$)\\ \hline
Twitter & 14.6s (1$\times$) & 17.5s (1.19$\times$)\\ \hline
RMAT 25 & 7.08s (1$\times$) & 11.1s (1.56$\times$)\\ \hline
RMAT 27 & 21.9s (1$\times$) & 42.8s (1.95$\times$)\\ \hline
SD & 15.0(1$\times$) & 19.7 (1.31$\times$) \\ \hline
\end{tabular}
\caption{Between Centrality runtime for 12 different starting points comparisons with Ligra and slowdown relative to Cagra}
\label{table:BCComparisons}
\end{table}

\subsection{Comparison with Hand Optimized Implementations and Other Frameworks}
\label{sec:compframeworks}

Tables~\ref{table:PageRankComparisons} to~\ref{table:BCComparisons} compare the running time for \Cagra with that of GraphMat, Ligra and GridGraph.

\punt{
\begin{table}[!h]
\center
\tabcolsep 5pt
\footnotesize
\begin{tabular}{p{1.5cm}|r|r}
\hline
Dataset & Optimized Version & Ligra (Baseline) \\ \hline 
LiveJournal  & 0.36s (1$\times$) & 0.33s (0.93$\times$)\\ \hline
Twitter & 2.91s (1$\times$) & 3.18s (1.09$\times$) \\ \hline
RMAT 25 & 1.14s (1$\times$) & 1.42s (1.24$\times$) \\ \hline
RMAT 27 & 4.53s (1$\times$) & 7.02s (1.54$\times$) \\ \hline
SD & 9.08s (1$\times$) & 10.8s (1.18$\times$) \\ \hline
\end{tabular}
\caption{BFS runtime for 12 different starting points comparisons with
  Ligra and slowdown (against each as a
  baseline). }
\label{table:BFSComparisons}
\end{table}
}

\punt{
\begin{table}[t]
\center
\tabcolsep 5pt
\footnotesize
\begin{tabular}{r|r|r}
\hline
Frameworks & Running Time & Slow Down  \\ \hline 
GridGraph  & 12.86s	& 5.04$\times$	\\ \hline
X-Stream &  18.22s	& 7.1$\times$ 	\\ \hline
GraphMat  &  4.20s	& 1.64$\times$  \\ \hline
Cagra & 2.55s & 1.00$\times$ \\ \hline
\end{tabular}
\caption{Execution time for 20 iterations of in-memory PageRank on the LiveJournal Graph on i2.xlarge with 4 cores}
\label{table:comparediskbasedframeworks}
\end{table}

\punt{
\begin{figure}[t]
	\includegraphics[width=0.5\textwidth]{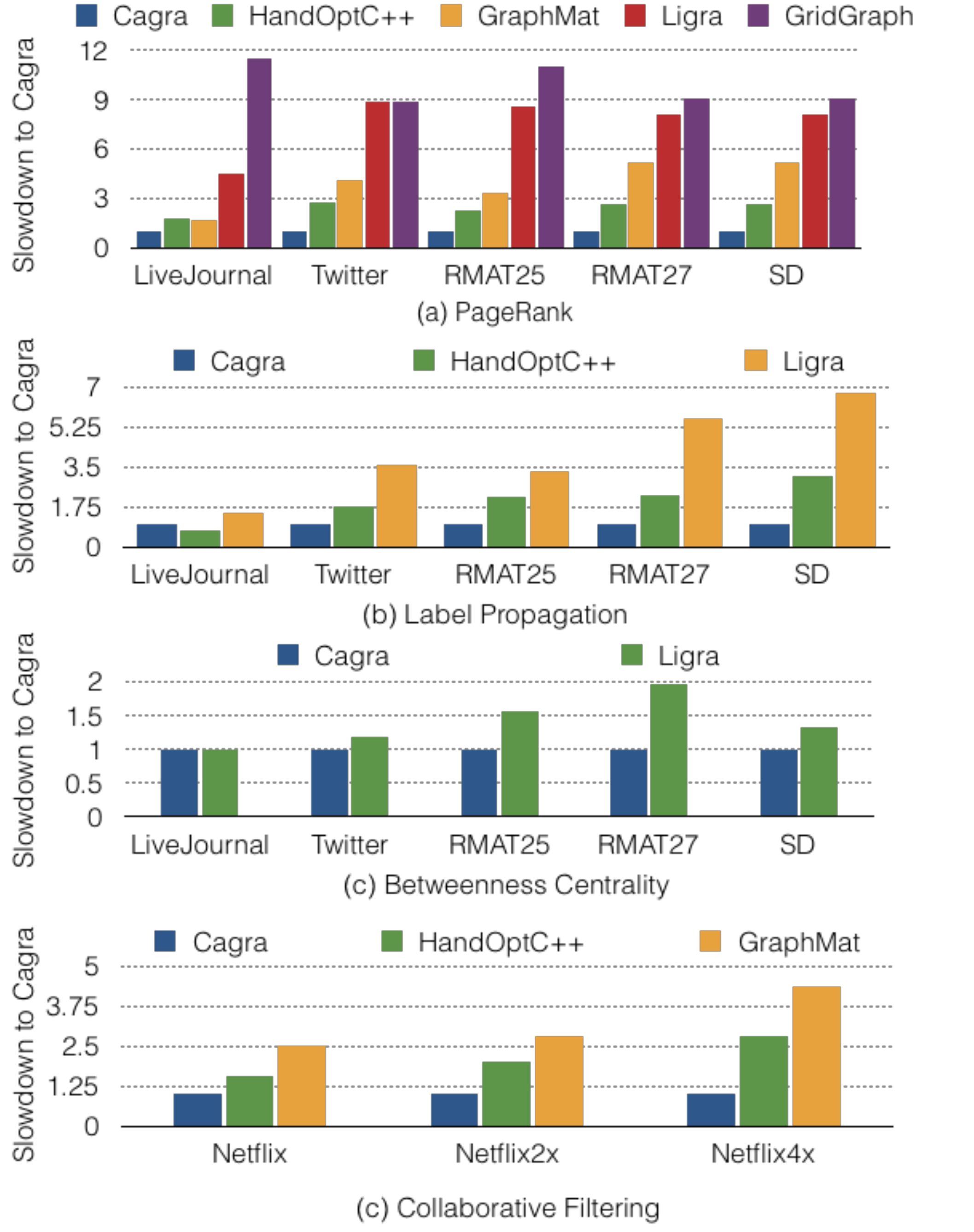}
	\caption{Slowdown of other frameworks relative to Cagra}
	\label{fig:Cagra_others}
\end{figure}
}
}

\textbf{Hand Optimized C++ Implementations}: We used hand optimized C++ implementations for PageRank, Label Propagation and Collaborative Filtering. These implementations are based on previous work~\cite{Satish14Sigmod}, and included many optimizations, such as hand-tuned work stealing load balance scheme, replacing expensive divisions to multiplications of reciprocal, working set compression by precomputing rank divided by degree, vectorization of loads, software prefetching and removal of unnecessary branches. These implementations do not apply \segmenting and \reordering.

\punt {
Our baselines for PageRank, Label Propagation and Collaborative Filtering are hand coded implementations that included many state-of-the-art optimizations, and are often faster than best results from existing frameworks. Our PageRank calculated the contribution (rank/degree) of each vertex beforehand to reduce the number of random accesses and divide instructions. 
	We use Ligra as our BFS and BC baseline to take full advantage of the push and pull switch optimization, missing from GraphMat and GridGraph. 
	}

\punt {
\textbf{\Cagra}: \Cagra applies cache optimizations to the applications. We compare our optimized versions to high-performance  to avoid exaggerating our performance gains due to the overheads in existing frameworks 
}

\textbf{Existing Frameworks}: GraphMat holds the record of the fastest published implementation of PageRank and Collaborative Filtering and is over 2x faster than Galois~\cite{nguyen13sosp-galois}. Collaborative Filtering is only implemented in GraphMat. GridGraph partitions the vertex data and the graph to improve cache performance. We used the number of partitions suggested in the GridGraph paper which we verified gave their best performance, since our machine has a similar LLC size. Ligra has the fastest implementations of Betweenness Centrality on many real world graphs thanks to its innovative push and pull switch optimization, and it is comparable to Galois on power law graphs.

\punt{
Table~\ref{table:comparediskbasedframeworks} compares \Cagra's cache optimized PageRank with GraphMat and the best running time reported for GridGraph~\cite{Zhu15ATC-GridGraph} and X-Stream~\cite{roy13sosp-xstream} on the LiveJournal graph. The experiments conducted on an i2.xlarge instance to show the portability of \Cagra across different hardware platforms. The whole graph fits in the memory of this system.}

Table~\ref{table:PageRankComparisons} to Table~\ref{table:LPComparisons} show that \Cagra's PageRank, Collaborative Filtering and Label Propagation's performance improves with the size of the graph. For PageRank, we only achieved 1.6x speedup on the LiveJournal graph compared to GraphMat because the graph is relatively small and most of the frequently referenced data fits in the last level cache. Betweenness Centrality has smaller working set than the other applications.

\begin{figure*}[t]
	\centering
	\includegraphics[width=0.7\textwidth]{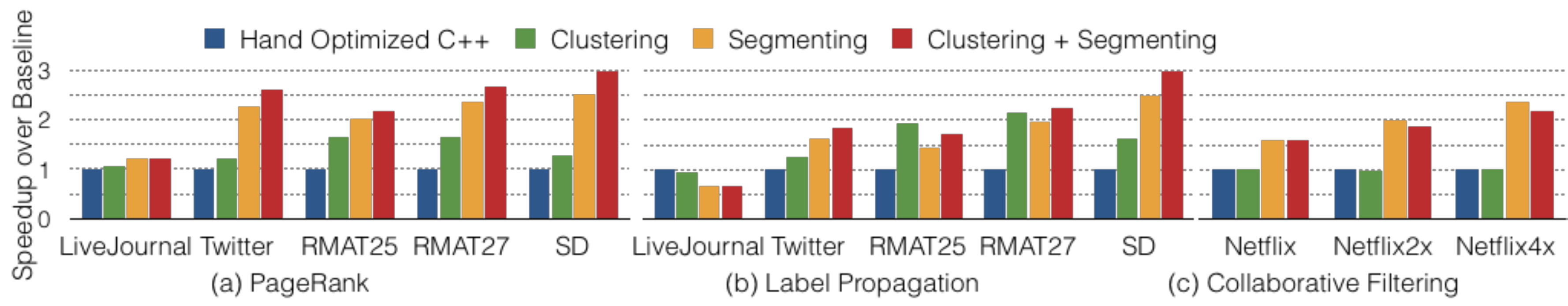}
	\caption{Speedups of optimizations on PageRank, Label Propagation, Collaborative Filtering}
	\label{fig:speedup}
\end{figure*}

\begin{figure*}[t]
	\centering
	\includegraphics[width=0.7\textwidth]{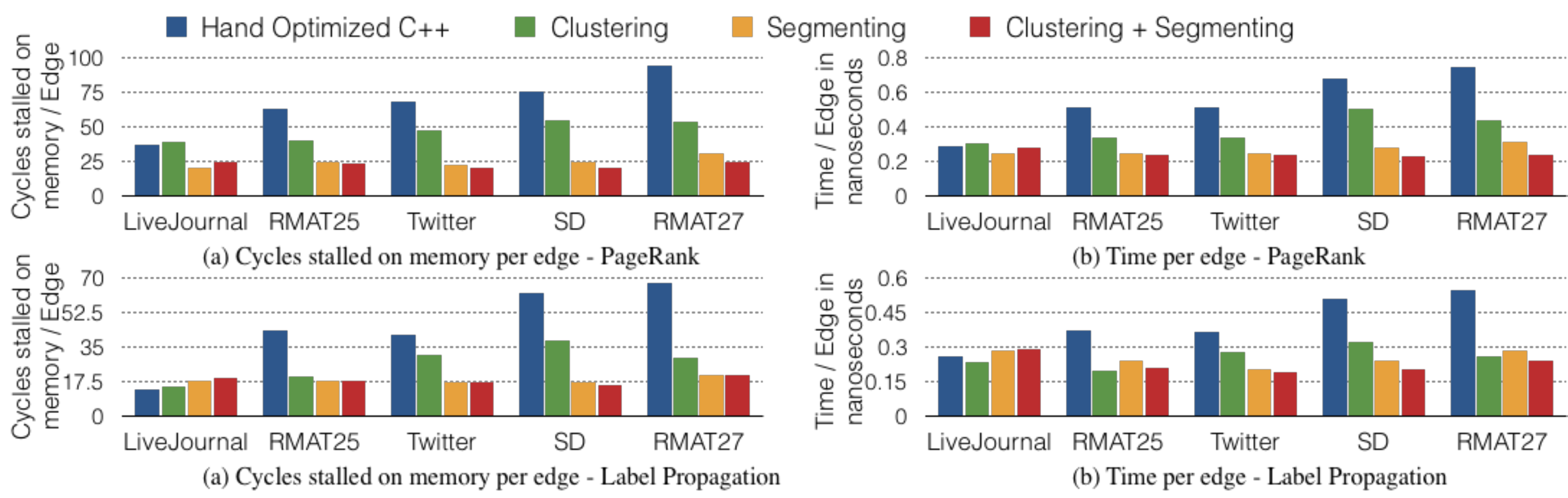}
	\caption{Cycles stalled on memory and time per edge for PageRank and Label Propagation. Cycles stalled per edge for Clustering + Segmenting is low and stable across graphs with increasing sizes, demonstrating that random accesses are confined in LLC.}
	\label{fig:stall}
\end{figure*}

\punt{ 
Table~\ref{table:PageRankComparisons} to Table~\ref{table:LPComparisons} show that our optimized PageRank, Collaborative Filtering and Label Propagation's performance improves with the size of the graph. For PageRank, we only achieved 1.6x speedup on the LiveJournal graph compared to GraphMat because the graph is relatively small and most of the frequently referenced data fits in the last level cache. However, for RMAT27, the graph is large enough that the frequently accessed data sets cannot be stored in cache without clustering. As a result, optimizing locality with \reordering and \segmenting\ achieves 4.30x speedup.

Table~\ref{table:BCComparisons} and Table~\ref{table:BFSComparisons} show that the speedups of our optimized BFS and BC implementations improves as the graph sizes increase. For the same reason, we are not seeing a significant speed up on the LiveJournal graph, because the frequently referenced nodes fit in the last level cache. Apart from being a larger graph, we are better on the RMAT27 graph compared to the Twitter and SD graph because the original graph is already ordered in a way that groups together neighbors while RMAT27 has a random ordering.
}

\subsection{Analysis of Optimizations}

In this section, we demonstrate the speedups of the optimizations for various applications in \figurename~\ref{fig:speedup} and show the reduction in cycles stalled on memory in \figurename~\ref{fig:stall}.

\textbf{\Segmenting}: Segmenting alone accelerates PageRank, Label Propagation and Collaborative Filtering by more than 2x. Segmented algorithm serves all of the random read requests in LLC, eliminating random DRAM access. The impact of segmenting on time and cycles stalled on memory per edge is evident in \figurename~\ref{fig:stall}.

	The cycles stalled on memory per edge for Segmenting stay low and stable for both PageRank and Label Propagation across graphs of increasing sizes. The stability comes from the fact that all random accesses are served in LLC, with almost a fixed latency. On the other hand, the \baseline and clustering's cycles stalled on memory per edge increases as we increase the size of the graph because more random reads are served in DRAM. We have also measured the LLC miss rate and find that it dropped from 46\% to 10\% on the Twitter graph after \segmenting\ has been applied. 
\punt{
The speedups with BC is not as significant because the working set is smaller than other applications. The vertex data for BC already mostly fits in LLC, with few random access to DRAM. 

The speedups with BC is not as significant because the working set is not large enough to benefit from \segmenting\. The vertex data for BC already mostly fits in LLC, with few random access to DRAM. 
}
\punt { 
For BC, at the evaluated graph sizes the effective working set of
application per-vertex data is well captured by caches.  For example,
for BC on the original Twitter42m graph started from the first userID,
the longest iteration activates 33\% of destination vertices, while
only 4\% of the source vertices are active.  While a bitvector check
is needed to check all vertices (5MB total), only 4\% of the vertices
will have to also bring in a \texttt{double} (13MB useful data) with
poor spatial locality but good temporal locality.  
}

\textbf{\Reordering}: Clustering is effective on PageRank, Label Propagation, Betweenness Centrality. \figurename~\ref{fig:stall} demonstrates that clustering significantly reduces cycles stalled. For Betweenness Centrality, clustering can help take advantage of the higher level L2 caches as the working set is not much larger than LLC. On Collaborative Filtering, full cache lines are used for per-vertex latent factor vectors, leaving little room for cache line utilization improvements.

\punt{
\textbf{\Reordering}: Clustering is effective on PageRank, Label Propagation, Betweenness Centrality. \figurename~\ref{fig:stall} demonstrates that clustering significantly reduces cycles stalled. For Betweenness Centrality, most of \Cagra's speedups comes from \reordering as the working set is too small for \segmenting to be effective. On Collaborative Filtering full cache lines are used for per-vertex latent factor vectors, leaving little room for cache line utilization improvements. 
}

\punt{ 
For BFS and BC, using bitvector to keep track of the active vertices set is another cache optimization many frameworks adopt for improved performance~\cite{Satish14Sigmod, sundaram15vldb-graphmat}. We implemented this optimization in Ligra to compare with \reordering. Figure~\ref{fig:speedup} shows that clustering can be as good as bitvector compression without modifying the Ligra framework. Combining clustering with bitvector, we gain an additional 20 percent speedup. Clustering is less effective for LiveJournal and Twitter because they are already in BFS based order that matches these access patterns.}

\textbf{Combining \Reordering and \Segmenting}: Combining the two techniques achieved even better results on PageRank and Label Propagation because clustering can further make better use of L2 cache within each segment that fit in LLC. The combined technique can further reduce 10-20\% of cycles stalled on memory, achieving another 20\% speedup over segmenting alone on large graphs, including RMAT27 and SD. 

\punt {Figure~\ref{fig:speedup-PageRank} shows that the optimizations stack on top of each other to improve the overall performance of PageRank. Frequency based clustering is most effective on RMAT27 because RMAT27 came with a random vertices order and both LiveJournal and Twitter graph came with a neighbor grouping order similar to BFS. This BFS order preserves some community structure in the graph, improving its locality. }

\subsection{Comparison with Other Cache Optimizations}
\label{sec:eval-orderings}

In this section, we compare \Cagra with Hilbert Ordering and GridGraph. Several researchers~\cite{yzelman14tpds-spmv,mcsherry15hotos} have
proposed traversing graph edges along a Hilbert curve, creating locality in both the source vertex read from and the destination vertex written to.
 
On a single core, processing the edge list in Hilbert order 
matches the serial performance of segmenting with clustering.
On multiple cores, however, we found that the technique did not
scale as well as our approaches.

\punt{
Our version using atomic updates for the writes
and at 1.2s per iteration, matches GraphMat framework, but is 
slower than our pull-based baseline.  
}

We tested two approaches to parallelize Hilbert-ordered updates.
The first, labeled HAtomic, uses atomic compare-and-swap updates.  While this approach scales
linearly with the number of threads, performance of atomic operations
is 3$\times$ worse than non-atomic operations.
The second, HMerge, uses an approach from \cite{yzelman12ecmi-cache-oblivious-spvm}
that creates per-thread private vectors to write updates to, and merges them at the end.
Only 5\% of the runtime is spent on merging the private vectors.  

\begin{figure}[t]
 \centering
 \includegraphics[width=0.6\columnwidth]
{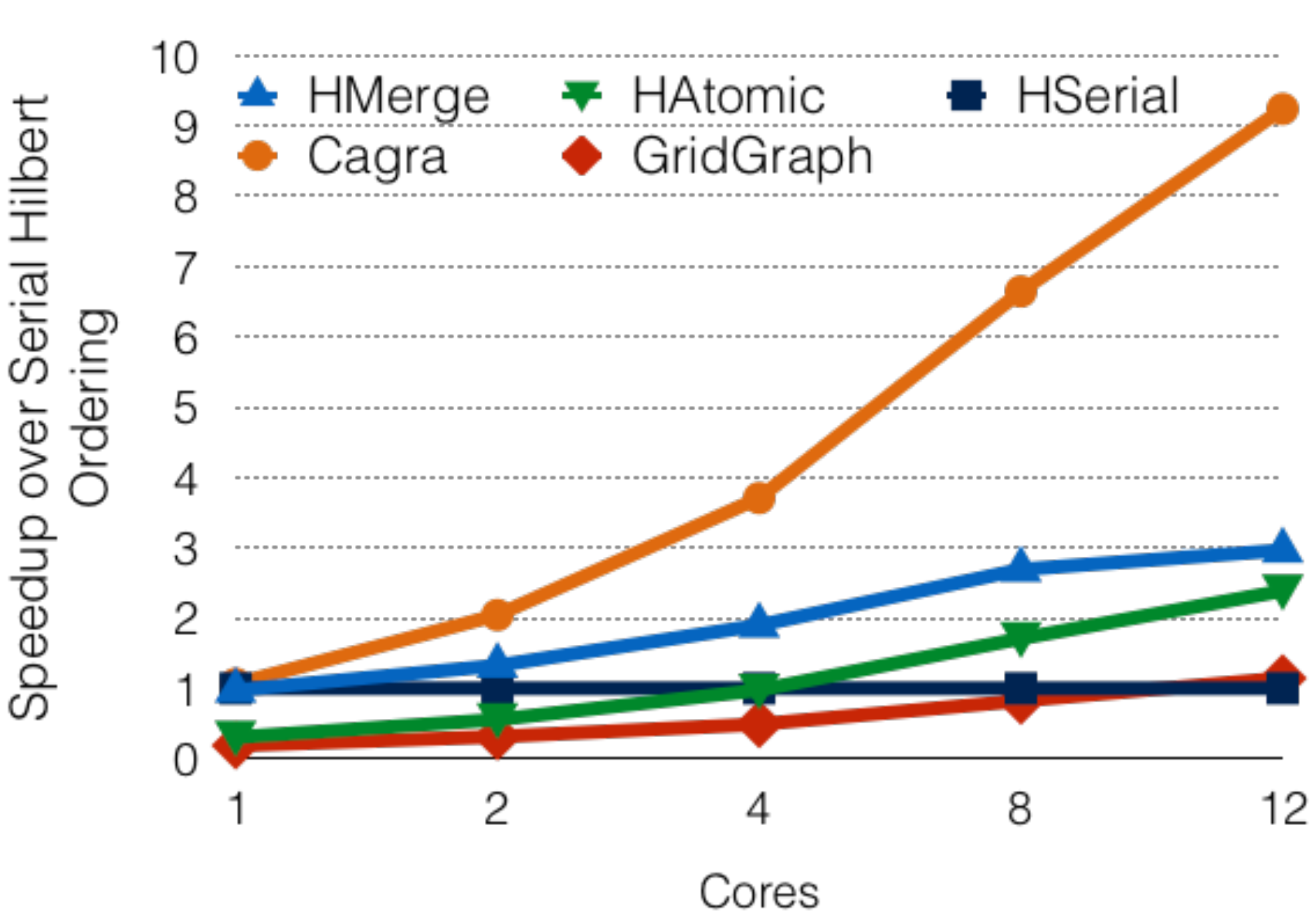}
 \caption{PageRank speedups on Twitter graph over HSerial (serial Hilbert-order) of 
   HAtomic (parallel Hilbert ordering with atomics), 
   HMerge (parallel Hilbert ordering with buffers), 
   \Cagra and GridGraph.}
 \label{fig:hilbert-speedup}
\end{figure}

Figure~\ref{fig:hilbert-speedup} shows the scalability on
PageRank of parallel Hilbert-order implementations
using a single NUMA socket.  
When using all 12 cores, the best runtimes of HSerial (5.4s),
HAtomic (2.3s), and HMerge (1.8s)
are ~3$\times$ slower than \Cagra, which takes 0.5 seconds. The main reason that Hilbert ordering does not scale well is cache contention.
Each core has a private L2 cache, however, the Last Level Cache is
competitively shared.  While Hilbert ordering helps increase locality
for each thread, because the threads work on independent regions, they
compete for the LLC. In contrast, segmenting allows multiple threads to
share the \emph{same} working set in the LLC, and continues scaling with
more cores.

\punt{
Peak performance for HMerge is reached with 10
cores, likely limited by the 20-way cache associativity of the LLC,
since each worker thread needs to access both source and destination
vector lines. 
}

\punt {
\begin{table}[htbp]
\center
\tabcolsep 5pt
\footnotesize
\begin{tabular}{rrrrrr}
\hline
HSerial & HAtomic & HMerge & HRacy* & Clustering & Segmenting \\ \hline 
5.4s    & XXX   s & 1.3s    & 0.9s   &  0.6s & 0.3s \\ \hline
\end{tabular}
\caption{PageRank/iteration in seconds (and speedups) on Hilbert vs
  Clustering on Twitter graph, using 24 cores (or 48 threads)}
\label{table:PageRankComparisons}
\end{table}

\begin{figure}[!t]
 \centering
 \includegraphics[width=0.6\columnwidth]
{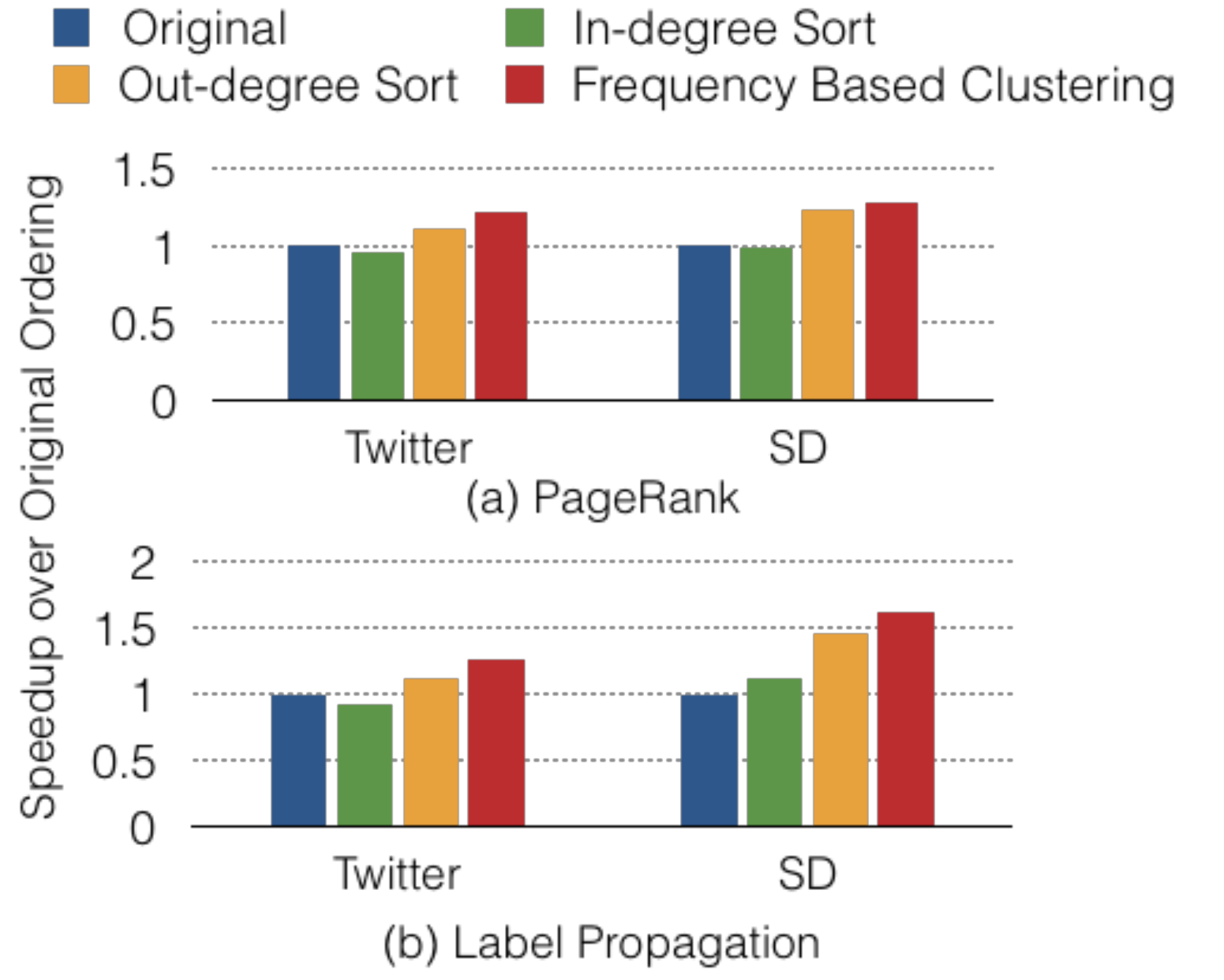}
 \caption{Comparison with other Orderings }
 \label{fig:ordering-speedup}
\end{figure}

Figure~\ref{fig:ordering-speedup} shows that, for real world graphs Twitter and SD, in-degree sorting is not as effective as out-degree based techniques. Frequency based clustering achieves better performance than out-degree sorting alone by preserving some locality in the original ordering of average degree vertices. 
}

\begin{figure}[t]
 \centering
 \includegraphics[width=0.6\columnwidth]
{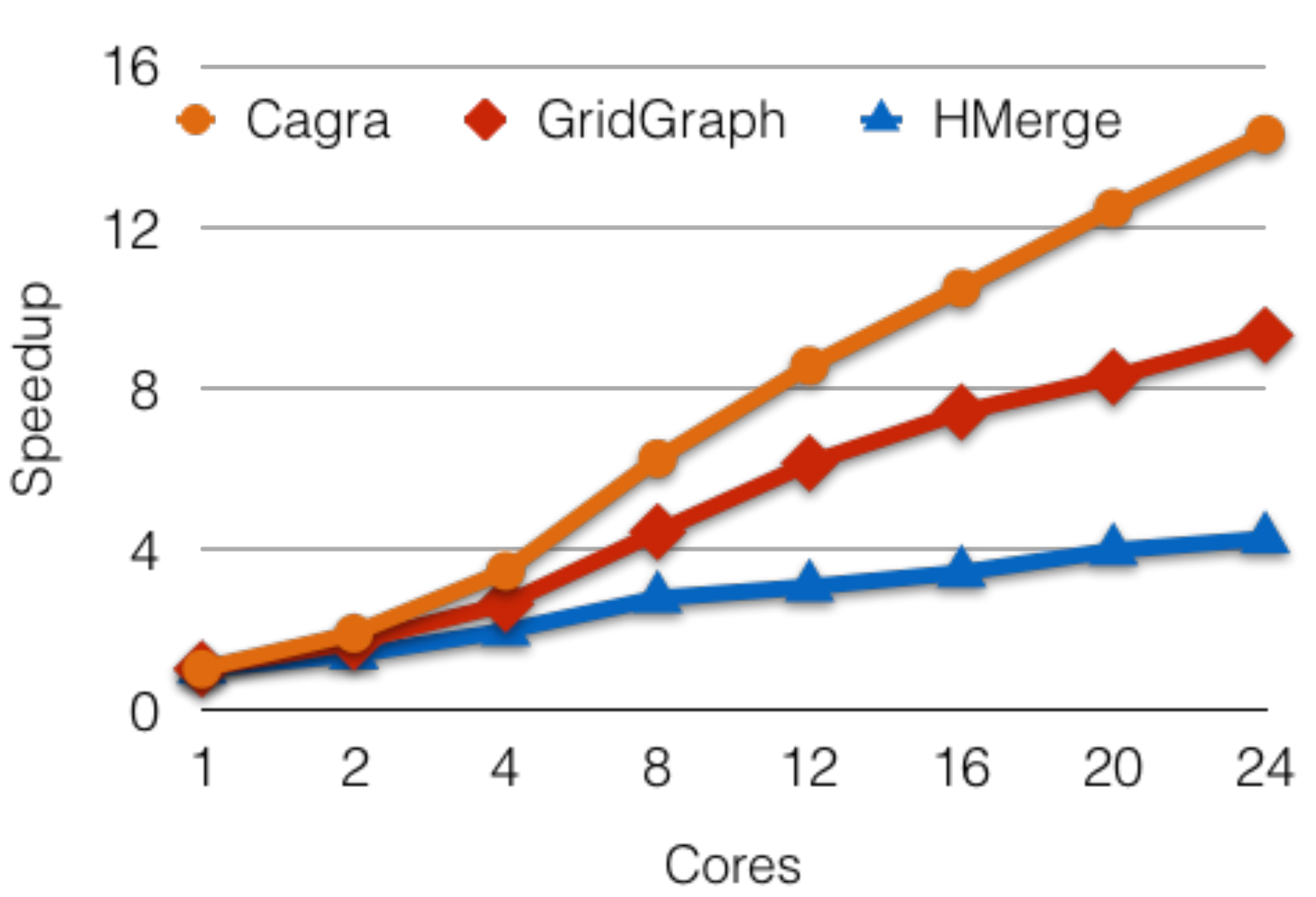}
 \caption{Scalability for PageRank on Twitter}
 \label{fig:pr-spedup}
\end{figure}

\figurename~\ref{fig:pr-spedup} shows that \Cagra is significantly more scalable than cache optimized GridGraph or parallel Hilbert ordering and achieves 8.5$\times$ speedup using 12 cores on the same NUMA socket, 14$\times$ speedup with 24 cores interleaved across two sockets, and 16$\times$ speedup with all 48 SMT using both hyperthreads per core. 

\punt{ 

\subsection{Scalability}

\begin{figure}[tbhp]
 \centering
 \includegraphics[page=1,trim=0cm 12cm 6cm 0cm, clip, width=4in]
{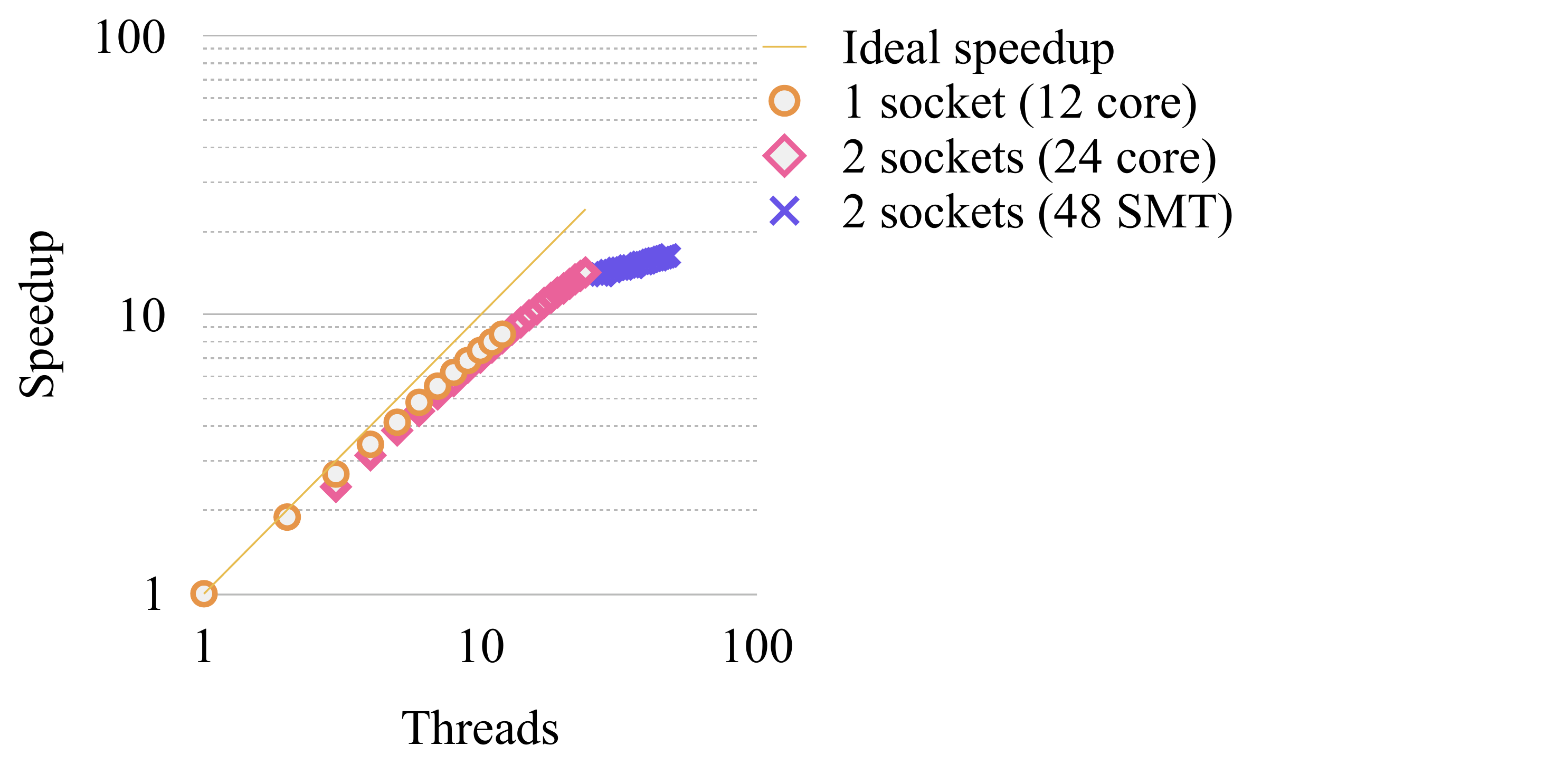}
 \caption{Speedups for PageRank on Twitter (original)
with 1 socket, 2 sockets, and using all SMT threads on 2 sockets}
 \label{fig:pr-spedup}
\end{figure}

\figurename~\ref{fig:pr-spedup} shows the scalability of PageRank in more detail compared to our
sequential code. We observe 8.5$\times$ speedup using 12 cores on the same NUMA socket,
14$\times$ speedup with 24 cores interleaved across two sockets,
and 16$\times$ speedup with all 48 SMT using both hyperthreads per core.
}

\subsection{Preprocessing Time}
\label{sec:preprocessingtime}

\begin{table}[ht]
\center
\tabcolsep 5pt
\footnotesize
\begin{tabular}{l|p{1.2cm}|p{1.2cm}|p{1.1cm}}
\hline
Dataset &  Clustering & Segmenting & Build CSR \\ \hline 
LiveJournal  & 0.1 s & 0.2 s & 0.48 s\\ 
Twitter & 0.5 s & 3.8 s & 12.7 s\\
RMAT 27 & 1.4 s & 6.3 s & 39.3 s\\ \hline
\end{tabular}
\caption{Preprocessing Runtime in Seconds.}
\label{table:PreprocessingOverheads}
\end{table}

Table~\ref{table:PreprocessingOverheads} shows \Cagra's preprocessing cost. \Cagra first computes the degrees of vertices and uses a parallel prefix sum to construct the CSR of the input graph. The CSR is then processed with clustering, and later partitioned into compressed subgraphs during segmenting. 

\Cagra's preprocessing cost is small compared to its significant performance gains. We do not include the preprocessing cost in Table~\ref{table:PageRankComparisons} to Table~\ref{table:BCComparisons}, since other frameworks also incur significant preprocessing costs that were not included. GraphMat and Ligra require the construction of CSR. \Cagra's clustering and segmenting cost can easily be amortized by performance gains over Ligra and GraphMat in 2--5 iterations, when applications, such as PageRank, can run for more than 20 iterations. GridGraph uses a special 2D grid representation of the graph that does not sort the edges by their destinations. However, 2D partitioning also takes significant more time than 1D partitioning used in \Cagra. GridGraph's preprocessing time for Twitter graph is 130s, much more than the 17s needed by \Cagra and is 9--11$\times$ slower than \Cagra on PageRank. Finally, the 1D segmented graphs can also be reused across multiple graph applications, further amortizing the preprocessing cost.

%% file: related.tex
\section{Related Work}
\label{sec:related}

There have been many projects optimizing graph computations in shared-memory systems, including Ligra, Galois, GraphMat and others~\cite{shun13ppopp-ligra, nguyen13sosp-galois, sundaram15vldb-graphmat}.
Satish et al.~\cite{Satish14Sigmod} benchmarked many of these frameworks and found them to underperform hand-optimized code.
The same authors proposed GraphMat~\cite{sundaram15vldb-graphmat}, a
framework based on sparse matrix operations that matched their
hand-optimized benchmarks.
Nonetheless, GraphMat still uses memory bandwidth inefficiently and does not optimize for cache aggressively.

\punt {so our optimizations (vertex reordering and segmenting) improve performance by up to 4$\times$ by improving cache line utilization and decreasing random access.}

\begin{table}[t]
\center
\footnotesize
\begin{tabular}{p{1.6cm}|p{1.6cm}|p{1.4cm}|p{1.6cm}}
\hline
Frameworks & \Cagra & GridGraph & X-Stream \\ \hline 
Partitioned Graph   & 1D-segmented CSR	& 2D Grid	& Streaming Partitions	 \\ \hline
Sequential DRAM traffic &E + (2q+1)V 	& E + (P+2)V 	&3E + KV \\ \hline
Random DRAM traffic  &0 	&0	& shuffle(E) \\ \hline
Parallelism & within 1D-segmented subgraph & within 2D-partitioned subgraph & across many streaming partitions \\ \hline
Runtime Overhead & Cache-aware merge 	&E*atomics	& shuffle and gather phase \\ \hline
\end{tabular}
\caption{Comparisons with other frameworks optimized for cache. E is
  the number of edges, V is the number of vertices, $q$ is the
  expansion factor for our techniques, P is the number of partitions
  for GridGraph, K is the expansion factor for X-Stream. 
  On Twitter graph, $E=36V, q=2.3, P=32$.}
\label{table:compframeworks}
\end{table}

 GridGraph~\cite{Zhu15ATC-GridGraph} and
 X-Stream~\cite{roy13sosp-xstream} claimed their techniques for reducing random 
 disk access can also be applied to reducing random memory
 access. GridGraph partitions the edges on both sources and destinations (2D partitioning) into subgraphs. Each subgraph is processed separately to make sure reads from sources and writes to destinations happen in cache. The issue with 2D partitioning is that the number of edges in each subgraph can be relatively small, making it difficult to scale efficiently beyond 4-6 cores when each subgraph is processed in parallel. \Cagra produces much better parallelism with a 1D segmented CSR scheme. Additionally, GridGraph uses atomics for parallel execution, incurring significant runtime overhead. X-Stream uses streaming partitions to keep the random reads in fast storage and reduce random writes in slow storage through a scatter-shuffle-gather design. X-Stream incurs heavy overhead by requiring additional sequential memory traffic for streaming the updates, extra random memory accesses and execution time for shuffle and gather phases. \Cagra completely eliminates random writes to DRAM and keeps the runtime overhead low with the cache-aware merge algorithm. Table~\ref{table:compframeworks} shows a detailed comparison between \Cagra, GridGraph and X-Stream. Techniques from other systems optimizing on the disk to memory boundary will also unlikely translate to performance gains as cache ptimizations. GraphChi~\cite{kyrola12osdi-graphchi} uses shards with Parallel Sliding Windows to keep random access low. However, it needs to stream the edges twice, and the updates from processing each interval will incur random writes in slow storage.  
 
 \punt{
 Additionally, X-Stream's approach of exploiting parallelism across multiple partitions makes inefficient use of the shared last level cache.

 FlashGraph~\cite{zheng15fast-flashgraph} stores a sorted edge list and partitions the graph in 2D, resulting in similar scalability issues.}
 
 	Polymer~\cite{zhang15ppopp-numa-polymer} is a NUMA-optimized framework
that focuses on minimizing both remote random access and cross-NUMA-node access. It uses a 1D graph partitioning scheme, but incurs slow remote writes for vertex data updates. While we do not focus on NUMA in this work, we believe that our techniques
could further improve intra-socket performance in Polymer.

Graph analytics has also been studied extensively in distributed memory systems~\cite{low12vldb-distr-graphlab, chen15eurosys-powerlyra}. These systems partition the graph into subgraphs that can be executed in parallel. Their partitioning model optimize for minimum communication and good load balance across different partitions. In contrast, \segmenting\ processes one segment at a time and optimizes for limiting the range of random access, instead of load balance. PowerLyra also exploits the skewed degree of the graphs with differentiated processing. 

\punt {
Graph analytics has also been studied extensively in distributed memory systems like Pregel, GraphLab and PowerLyra~\cite{malewicz10sigmod-pregel, low12vldb-distr-graphlab, chen15eurosys-powerlyra}. These systems partition the graph into subgraphs that can be executed in parallel. Their partitioning model is very different from segmenting. Distributed memory systems optimize for minimum communication and good load balance across different partitions that are expected to execute in parallel. In contrast, \segmenting\ processes one segment at a time and optimizes for limiting the range of random access, instead of load balance. PowerLyra also exploits the skewed degree of the graphs, similar to clustering, with differentiated processing in the distributed setting. 

}

Recent works have looked at speeding up graph application with vertex and edge reordering. A concurrent work~\cite{Wei2016-speeduporder} applied in-degree sort to many sequential graph algorithms. \Reordering improves the performance compared to in-degree sort by preserving locality in the original ordering of the graphs. Hilbert ordering~\cite{mcsherry15hotos} is an edge ordering technique that was shown to improve the cache performance of graph algorithms. We studied Hilbert ordering extensively in Section~\ref{sec:eval-orderings} and found that it underperforms our techniques on multicore systems.

For graphs with poor locality, such as uniform random graph, propagation blocking~\cite{beamer17ipdps} and milk compiler~\cite{Kiriansky16milk} achieves good cache performance by reorganizing random memory accesses into sequential ones at runtime. Cagra does more in preprocessing with little runtime overhead.

\punt{ 
Recent works have looked at speeding up graph application with vertex and edge reordering. A concurrent work~\cite{Wei2016-speeduporder} applied in-degree sort to many sequential graph algorithms. \Reordering achieves much better performance compared to in-degree sort by preserving locality in the original ordering of the graphs. Other techniques studied in the paper incur hard-to-amortize preprocessing overhead, up to 1.5 hrs for Twitter graph. We focus on lightweight techniques with low preprocessing overhead. Degree based reordering has also been used for reducing algorithmic complexity for Triangular Counting~\cite{jshun15ICDE}, while we focus only on the cache performance improvement. Hilbert ordering~\cite{mcsherry15hotos} is an edge ordering technique that was shown to improve the cache performance of single threaded PageRank. We studied Hilbert ordering extensively in Section~\ref{sec:eval-orderings} and found that it underperforms our techniques on multicore systems.
}

Finally, graph applications like PageRank are analogous to sparse
matrix-vector multiply problems, for which techniques, such as cache blocking, have been proposed~\cite{yzelman14tpds-spmv,williams-spmv}. \Segmenting performs better than previous cache blocking techniques as we do not fit both the sources and destinations (2D blocking) in cache. Our technique fit only the sources in cache (1D segmenting) and store the writes sequentially in large buffers, which are later processed using cache-aware merge. This approach allows us to generate greater parallelism, reduce preprocessing time and keep runtime overhead low. Additionally, not all applications, such as collaborative filtering, can be easily expressed as SpMv problems.

%% file: conclusion.tex
\section{Conclusion}
\label{sec:conclusion}

Graph analytics are an essential part of modern data analysis workflows,
leading to significant work to optimize them on shared-memory machines.
Graph applications inherently appear to have poor cache utilization, requiring a large number of random DRAM accesses. In this paper, we show that substantial performance improvements can be obtained by eliminating random DRAM access with \segmenting. \Segmenting uses a novel 1D segmentation and cache-aware merge scheme to achieve scalable performance with low runtime overhead. We then present our framework, \Cagra, that applies \segmenting to various graph applications with an easy to use interface. \Cagra is up to 5$\times$ faster than the best published results for common graph applications from high performance graph frameworks and up to 3$\times$ faster than expert optimized C++ implementations.

\punt{ 
The techniques should be broadly applicable to current graph
frameworks and parallelize well on multicores, significantly reducing cycles stalled on memory.
}


%% file: ack.tex
\section{Acknowledgments}

We thank William Hasenplaugh for discussions and suggestions on the paper, Julian Shun, MIT COMMIT group members and our reviewers for the many feedback. This research is supported by affiliate members and other supporters of the Stanford DAWN project (Intel, Microsoft, Teradata, and VMware) as well as NSF CAREER grant CNS-1651570, grant from Toyota Research Institute, DARPA grant FA8750-17-2-0126, DOE awards DE-SC008923 and DE-SC014204.